\documentclass[12pt,a4paper]{article}
\usepackage{epsfig,amssymb,euscript}
\usepackage{amsmath}
\usepackage{bbm}
\usepackage[linktocpage]{hyperref}

\def\be{\begin{equation}}
\def\ee{\end{equation}}
\newcommand{\eq}[1]{\begin{equation}#1\end{equation}}
\def\bea{\begin{eqnarray}}
\def\eea{\end{eqnarray}}
\newcommand{\ul}{\underline}

\numberwithin{equation}{section} 
\usepackage[]{graphicx}
\def\cala         {{\cal A}}

\def\calb         {{\cal B}}
\def\calc         {{\cal C}}
\def\cald         {{\cal D}}

\def\calf         {{\cal F}}

\def\cali         {{\cal I}}
\def\calj         {{\cal J}}
\def\calk         {{\cal K}}
\def\call         {{\cal L}}

\def\caln         {{\cal N}}
\def\calo         {{\cal O}}

\def\cals         {{\cal S}}

\def\calv         {{\cal V}}
\def\calw         {{\cal W}}
\def\calz         {{\cal Z}}

\def\Re           {{\rm Re\hskip0.1em}}
\def\Im           {{\rm Im\hskip0.1em}}

\def\sqr#1#2{{\vcenter{\vbox{\hrule height.#2pt
 \hbox{\vrule width.#2pt height#1pt \kern#1pt \vrule width.#2pt}\hrule
 height.#2pt}}}}

\def\m{\mu}

\def\d{\text{d}}
\def\Z{{\cal Z}}
\def\T{{\cal T}}
\def\t{t}

\newcommand\bbone{\ensuremath{\mathbbm{1}}}

\def\slashchar#1{\setbox0=\hbox{$#1$}           
\dimen0=\wd0                                 
\setbox1=\hbox{/} \dimen1=\wd1               
\ifdim\dimen0>\dimen1                        
\rlap{\hbox to \dimen0{\hfil/\hfil}}      
#1                                        
\else                                        
\rlap{\hbox to \dimen1{\hfil$#1$\hfil}}   
/                                         
\fi}

\begin{document}
\font\cmss=cmss10 \font\cmsss=cmss10 at 7pt

\rightline{\small{\tt MPP-2007-92}}
\rightline{\small{\tt KUL-TF-07/12}}

\vskip .7 cm

\hfill
\begin{center}
{\Large \textbf{From ten to four and back again:\\
\vskip .3 cm
how to generalize the geometry
}}
\end{center}

\vspace{1cm}
\begin{center}
{\large\textsl{Paul Koerber~$^{a}$ and Luca Martucci~$^{b}$}}
\renewcommand{\thefootnote}{\arabic{footnote}}

\vspace{25pt}
\textit{\small $^a$ Max-Planck-Institut f\"{u}r Physik -- Theorie,\\
                    F\"{o}hringer Ring 6,  D-80805 M\"{u}nchen, Germany}\\ \vspace{6pt}
\textit{\small $^b$ Institute for Theoretical Physics, K.U. Leuven,\\ Celestijnenlaan 200D, B-3001 Leuven, Belgium}\\  \vspace{6pt}
\end{center}

\vspace{12pt}

\begin{center}
\textbf{Abstract}
\end{center}

\vspace{4pt} {\small  We discuss the four-dimensional $N=1$ effective approach in the study of warped type II flux compactifications
with $SU(3)\times SU(3)$-structure to AdS$_4$ or flat Minkowski space-time. The non-trivial warping makes it natural to use a supergravity formulation invariant under local complexified Weyl transformations. We obtain the classical superpotential from a standard argument involving domain walls and generalized calibrations and show how the resulting F-flatness and D-flatness equations exactly reproduce the full ten-dimensional supersymmetry equations.  Furthermore, we consider the effect of non-perturbative corrections to this superpotential arising from gaugino condensation or Euclidean D-brane instantons. For the latter we derive the supersymmetry conditions in $N=1$ flux vacua in full generality. We find that the non-perturbative corrections induce a quantum deformation of the internal generalized geometry.
Smeared instantons allow to understand KKLT-like AdS vacua from a ten-dimensional point of view. On the other hand, non-smeared instantons in IIB warped Calabi-Yau compactifications `destabilize' the Calabi-Yau complex structure into a genuine generalized complex one. This deformation gives a geometrical explanation of the non-trivial superpotential for mobile D3-branes induced by the non-perturbative corrections.

    \noindent }

\vspace{1cm}

\thispagestyle{empty}


\vfill
\vskip 3.mm
\hrule width 5.cm
\vskip 2.mm
{\small
\noindent e-mail: \texttt{koerber@mppmu.mpg.de}, \texttt{luca.martucci@fys.kuleuven.be}}

\newpage

\setcounter{footnote}{0}

\tableofcontents

\newpage

\section{Introduction}

The realization of physically relevant models has always been one of the main objectives of string theory. Therefore, configurations with the observable four dimensions in a distinguished role are of great interest. For various reasons, both phenomenological as for comparative ease of finding solutions,
supersymmetry in these four dimensions is favoured. Supersymmetry breaking is then to take place at a lower energy scale. Because of this, Calabi-Yau manifolds have always made for one of the more natural starting points for searching for interesting effective four-dimensional theories.

In the last decade it has been realized that the possibilities for realistic models can be enormously enlarged by adding fluxes, branes and orientifolds to the internal compactification space (for recent reviews on the subject and historically more complete lists of references, see \cite{granareview,morereviews}). There are two approaches to the problem, a four-dimensional and a ten-dimensional one. In the four-dimensional approach, one typically starts from the effective low-energy four-dimensional theory arising in an ordinary fluxless compactification and then adds the effect of fluxes as a modification of this four-dimensional theory: typically some of the scalars picking up a charge and a potential
being generated \cite{polchflux,gaugeflux}. In the second approach on the other hand, one tries to construct and study full ten-dimensional configurations with fluxes, which in the general case means that one has to leave the familiar realm of Calabi-Yau manifolds. The four-dimensional approach is obviously more practical for searching effective theories with the desirable physical properties. However, since the introduction of fluxes on the internal manifold can drastically change the geometrical and topological properties of the original vacuum, one is confronted with the problem to what extent the low-energy theory essentially based on the Calabi-Yau structure of the fluxless compactification can still be trusted, calling
for a better understanding of the relation between the four- and ten-dimensional approaches.

For this reason, it is desirable that the four-dimensional effective theories be based instead on the broader class of internal manifolds of $SU(3)\times SU(3)$-structure describing the most general geometric flux vacua. Indeed, it has been realized recently that minimally supersymmetric type II flux configurations are naturally described using the language of generalized complex geometry \cite{hitchin,gualtieri}. For example, the supersymmetric conditions for such general flux vacua can be elegantly written in terms of `polyforms' (sums of forms of different degree)  \cite{gmpt} which constitute the basic objects of generalized geometry. Gratifyingly, the same is true for D-branes in these
vacua, which are equipped with generalized calibrations \cite{gencalpaul,luca1,luca2,paulluca,lucajarah,lucanapoli} providing
a physical explanation of the polyform equations in \cite{gmpt}.

Using an approach similar to the ones of \cite{granaN2,grimm,granaN2bis} we clarify in this paper how these $N=1$ supersymmetric flux compactifications to both AdS or flat $\mathbb{R}^{1,3}$ space-time can be understood from a four-dimensional point of view. Two problems that hinder a full reduction are that the fluxes generically mix different Kaluza-Klein scales and the moduli space of the generalized compactifications is still not completely understood \cite{alessandro}. Therefore, it is not easy to identify the zero- or light modes and integrate out the `massive' modes
in order to obtain a four-dimensional supergravity with a finite number of fields.
We will not attempt such a reduction here -- the problem is studied in $N=1$ or $N=2$ contexts in e.g.~\cite{micu1,house,granaN2,kashanipoor,grimm,granaN2bis,micu2,mario1} -- but instead keep all Kaluza-Klein modes. This procedure should be considered as a first step to full reduction. However, it turns out that -- as we will explain in a moment -- we can already gain new insights when adding a non-perturbative superpotential, a quantity that is usually associated to the four-dimensional description.

A related complication is the presence of a non-trivial warp factor multiplying the four-dimensional metric, which is often approximated to be constant. In effect, this means neglecting the back-reaction of fluxes, D-branes and orientifolds on the geometry.
Moreover, a non-trivial warp factor is a physically important feature in flux compactifications (see e.g.~\cite{gkp,giddings1}) and in flux vacua holographically dual to (possibly confining) gauge theories.  Thus, in this paper we will allow for a non-trivial warp factor, which
turns out to naturally lead to a four-dimensional $N=1$ effective description that is invariant under complexified Weyl transformations. Such a theory can be obtained by partially gauge-fixing the $N=1$ superconformal supergravity presented e.g.\ in \cite{supconf}. The resulting theory can be considered as morally in the string frame.
If needed, this formulation can always be `gauge-fixed' to the standard Einstein frame, making contact with previous results. For example our approach offers a natural way to treat the `warped' K\"ahler potentials proposed in \cite{giddings1,giddings2} in the context of IIB warped Calabi-Yau compactifications.

In this paper we do not directly derive the complete four-dimensional supergravity action,
but instead extract from simple arguments the ingredients needed to understand the structure of supersymmetric vacua. For example the superpotential with correct dependence on the warp factor will be derived by extending a standard argument \cite{gukov} involving (generalized) calibrations and domain walls to the generalized setting. In the constant warp factor approximation, our result for the superpotential reduces to those obtained in
\cite{granaN2,grimm,granaN2bis} by direct dimensional reduction. Our approach essentially still contains all the information on the ten-dimensional
theory, and as a check on the superpotential and (conformal) K\"ahler potential we will show that {\em all} the ten-dimensional supersymmetry conditions can be understood as D- and F-flatness conditions of our the four-dimensional theory.

Our approach also offers the possibility to add a non-perturbative correction to the superpotential and see its effects on the ten-dimensional
geometry. Indeed, non-perturbative effects arising from instantonic D-branes or strong-coupling effects on stacks of `confining' space-filling D-branes have played an important role in many models. For example they are used to stabilize some background moduli along the line of \cite{kklt} or, more recently, they can induce a non-trivial superpotential for mobile space-filling D3-branes that would classically be free to move in the internal space \cite{ganor,haack,klebanov}.  As an example of how our
four-to-ten approach allows to inspect the non-perturbative effects on the internal geometry of the compactification, we will show how the AdS vacua found in \cite{kklt} can be understood from a ten-dimensional point of view as arising from smeared instantons. On the other hand, and perhaps more importantly, a non-perturbative correction generically deforms the generalized (complex) structure of the classical vacuum. In particular, the most studied examples of IIB compactifications involve warped Calabi-Yau internal manifolds, which are thus complex in the ordinary sense. In this case the non-perturbative corrections are generated by Euclidean D3- or confining D7-branes and a simple argument will lead us to conclude that they destabilize the internal ordinary complex structure into a genuine {\em generalized} complex  one. This deformation generates a geometrical superpotential \cite{luca2} for the classically free space-filling D3-branes. We discuss how this geometrical superpotential is in perfect agreement with the ones computed in \cite{haack,klebanov} in completely different ways.

In order to lighten the presentation of the main results, many technical details have been relegated to appendices,
albeit they include also some new results. In particular we identify the proper holomorphic structure on the space of fluctuations of the $SU(3)\times SU(3)$-structure  flux compactifications, which can also be useful in a future study of the moduli space, and we derive the supersymmetry/calibration conditions for instantonic Euclidean D-branes on general $N=1$ flux vacua (which as expected are directly related to the space-filling D-branes considered in \cite{luca1,luca2,paulluca}). We also find that the supersymmetric instanton action depends holomorphically on as well the open as the closed string degrees of freedom, while the anti-instanton is anti-holomorphic. These results on instantons may also be useful in the gauge/gravity correspondence (see e.g.~\cite{klebanov2}).

In section \ref{back} we review the description of ten-dimensional $N=1$ flux compactifications in the language of generalized geometry, while
in section \ref{fourdim} we set up the four-dimensional description. Indeed, in the first two subsections we extract the superpotential
from a Gukov-Vafa-Witten type argument and identify the holomorphic variables, the fluctuations of which are studied in more detail
in appendix \ref{infdef}. We discuss briefly open string degrees of freedom in the next subsection, and put the ingredients of the four-dimensional
description together and extract the K\"ahler potential in the last subsection. More details on the Weyl invariant formalism we are using can be found in appendix \ref{sugrapp}. In section \ref{10susy} we show how to derive the complete set of ten-dimensional supersymmetry equations
as F- and D-flatness conditions, for which we provide the example of IIB warped Calabi-Yau compactifications in section \ref{oldsup}. We demonstrate
the effects of adding a non-perturbative correction to the superpotential on the internal geometry in section \ref{corrections} and provide the examples of KKLT-like AdS vacua and mobile D3-branes.
We end with conclusions. Apart from the already mentioned appendices, appendix \ref{hodge} reviews some generalized geometry lore and in
particular the generalized Hodge decomposition of forms. Appendix \ref{appinst} derives the conditions for obtaining a supersymmetric
D-brane instanton in a flux vacuum and shows that its action is holomorphic, when its deformations are dressed with the naturally induced
complex structure. Appendix \ref{orientifolds} describes some properties of supersymmetric orientifolds in flux compactifications.

\section{Ten-dimensional supersymmetric vacua}
\label{back}

We consider minimally supersymmetric type II vacua where the ten-dimensional space-time
takes a warped-factorized form $X\times_w M$, with $X$ either AdS$_4$ or flat $\mathbb{R}^{1,3}$, and $M$ the internal six-dimensional space.
Thus the ten-dimensional metric has the form
\bea\label{metric}
\d s^2=e^{2A(y)} g_{\mu\nu}(x)\d x^\m\d x^\nu + h_{mn}(y)\d y^m\d y^n \ ,
\eea
where $g$ is the metric on $X$ and $h$ is the metric on $M$. All the other supergravity fields also preserve this splitting and furthermore, since we require $N=1$ four-dimensional supersymmetry, our background admits four independent Killing spinors of the form
\bea\label{adskilling}
\epsilon_1 &=&\zeta_+\otimes \eta^{(1)}_+ \, + \, \zeta_-\otimes \eta^{(1)}_- \ ,\cr
\epsilon_2 &=&\zeta_+\otimes \eta^{(2)}_\mp \, + \, \zeta_-\otimes \eta^{(2)}_\pm \ ,
\eea
for IIA/IIB.\footnote{Here and in the following, we mean that the upper sign is for IIA while the lower is for IIB.} The Majorana conditions for $\epsilon_{1,2}$
imply the four- and six-dimensional reality conditions $(\zeta_+)^*=\zeta_-$ and $(\eta^{(1,2)}_+)^*=\eta^{(1,2)}_-$. We will always impose on the flux vacua the condition that the norms of the internal spinors be equal, $\eta^{(1)\dagger}_+\eta^{(1)}_+=\eta^{(2)\dagger}_+\eta^{(2)}_+=|a|^2$. This condition is necessary for supersymmetric AdS vacua \cite{lusttsimpis,scan} or for introducing supersymmetric D-branes or O-planes \cite{luca1}.
In (\ref{adskilling}) the two internal chiral spinors $\eta^{(1)}_+$ and $\eta^{(2)}_+$ are fixed for a specific solution. In fact, they define an $SU(3)\times SU(3)$-structure of $T_M\oplus T_M^\star$ and characterize the solution. On the other hand, $\zeta_+$ is any of the four independent  Killing spinors satisfying the equation
\begin{equation}
\nabla_\mu \zeta_-=\pm \frac12 W_0 \gamma_\mu \zeta_+ \label{spinori}\ ,
\end{equation}
for IIA/IIB. In the following section we will explain how $W_0$ is related to the on-shell value of the, properly normalized, superpotential of the
four-dimensional effective description, so that $|W_0|^2=-\Lambda/3$ where $\Lambda$ is the effective four-dimensional cosmological constant.

In \cite{gmpt} the supersymmetry conditions obtained from the Killing spinor equations were written in terms of polyforms obtained by the Clifford map from the bi-spinors $\eta^{(1)}_+\otimes\eta^{(2)\dagger}_\pm$.  It is convenient to introduce the normalized polyforms\footnote{Note that in \cite{luca2,paulluca,lucajarah,lucanapoli} the normalized pure spinors (\ref{nps}) were denoted by $\hat\Psi^{\pm}$,  while $\Psi^{\pm}$ referred to the ones via the Clifford map associated to $\eta^{(1)}_+\otimes\eta^{(2)\dagger}_\pm$ without normalization.}
\bea\label{nps}
\slashchar{\Psi}^{\pm} = -\frac{8i}{|a|^2} \, \eta^{(1)}_+\otimes\eta^{(2)\dagger}_\pm \ ,
\eea
and rename them as
\bea
\Psi_1=\Psi^{\mp}\quad\text{and}\quad \Psi_2=\Psi^{\pm}\qquad\text{in IIA/IIB.}
\eea

The polyforms $\Psi_1$ and $\Psi_2$ can viewed as spinors of $T_M \oplus T^\star_M$ and as such are pure. Furthermore they obey the following compatibility and normalization constraints
\begin{subequations}
\label{psconsts}
\begin{align}
\label{psconst2}
\langle \Psi_1,\mathbb{X}\cdot{\Psi}_2\rangle &=\langle \bar{\Psi}_1,\mathbb{X}\cdot{\Psi}_2\rangle=0 \, , \qquad \forall \mathbb{X}\in T_M\oplus T^\star_M\ ,\\
\label{psconst1}
\langle \Psi_1,\bar{\Psi}_1\rangle & =\langle \Psi_2,\bar{\Psi}_2\rangle=-8i\sqrt{\det h}\,\d^6y \, ,
\end{align}
\end{subequations}
where the Mukai pairing $\langle \cdot,\cdot \rangle$ is defined in \eqref{mukai}.
Each of the two globally defined pure spinors equips $M$ with an $SU(3,3)$-structure, or almost generalized Calabi-Yau structure. The conditions \eqref{psconsts} imply that these two structures combine into an $SU(3)\times SU(3)$-structure, which also provides a generalized metric $(h,B=0)$ for $M$. In this picture the degrees of freedom of the $B$-field sit in the field-strength $H$ instead.

By setting $W_0=e^{-i\theta}/R$, where $R$ is the AdS radius, the supersymmetry conditions found in \cite{gmpt} can be rewritten as the following two sets of equations
\begin{subequations}
\label{susyconds}
\begin{eqnarray}
\label{susycond1}
\d_H \big(e^{4A-\Phi} \Re \Psi_1 \big) & =& (3/ R) \, e^{3A-\Phi} \Re (e^{i\theta} \Psi_2) \mp e^{4A} \alpha(\star_6 F) \, ,\\
\d_H \big[e^{3A-\Phi}\Im (e^{i\theta}\Psi_2)\big]&=& (2/ R) \,e^{2A-\Phi}\Im \Psi_1\label{susycond2} \, ,
\end{eqnarray}
\end{subequations}
for IIA/IIB -- where $\alpha$ is the operator reversing the indices of a polyform -- and
\begin{subequations}
\label{intconds}
\bea\label{intcond1}
&&\d_H(e^{2A-\Phi}\Im \Psi_1)=0 \ ,\\
&&\label{intcond2}\d_H\big[e^{3A-\Phi} \Re (e^{i\theta} \Psi_2)\big]=0\ .
\eea
\end{subequations}
For the AdS case, where $W_0\neq 0$, the eqs.~(\ref{intconds}) follow as integrability conditions from (\ref{susyconds}),\footnote{We take into account the equations of motion for $F$.} while in the flat-space limit, $W_0\rightarrow 0$, they must be added as independent conditions.


\section{The off-shell four-dimensional description}
\label{fourdim}

In the previous section we have recapitulated the conditions for obtaining a minimally supersymmetric vacuum compactification of type II from a ten-dimensional point of view.
In the next section, we will show that it is possible to understand {\em all} of those equations
in terms of effective four-dimensional $N=1$ supergravity.
But first we must set up this four-dimensional description, which is the purpose of this section. We remark again that we are not making any truncation to light modes; in other words we are keeping all KK-modes.

In realistic flux compactifications, especially to flat $\mathbb{R}^{1,3}$ space-time, consistency often requires the introduction of orientifolds, which have been studied in this setting in \cite{grimm,scan,dimi}. We will assume them to be present when needed and in this case we will implicitly work on the covering space, where their effect can be seen as projecting out part of the spectrum, as reviewed in appendix \ref{orientifolds}. Using the results of that appendix, one can easily check that the following results are consistent with such a quotient.

\subsection{On-shell superpotential from domain walls}
\label{onshell}

Let us now derive the form of the superpotential from a simple argument involving domain walls, which was also used in \cite{gukov} in the specific case of Calabi-Yau compactifications.  Here, we will start with an $N=1$ generalized compactification with fluxes already present, and then probe it with D-brane domain walls and instantons preserving half of the supersymmetry. This will allow us to obtain a more general result, not necessarily linked to an underlying Calabi-Yau geometry. Our argument will give the properly normalized superpotentials, automatically including a possibly non-trivial warp factor. The result will be consistent with the superpotentials in the constant warp factor approximation, obtained in \cite{granaN2,grimm,granaN2bis} by dimensional reduction.

For simplicity, let us consider compactifications to flat $\mathbb{R}^{1,3}$. The result is however, as we will see, valid for AdS compactifications as well and the following analysis is readily extendable to AdS backgrounds by using the results of \cite{adsbranes}. Supersymmetric D-branes on such general flux compactifications have been studied in \cite{luca1,luca2,lucajarah}, where it was shown that the background is naturally equipped with generalized calibrations corresponding to the pure spinors describing the supersymmetry. In particular, $e^{3A-\Phi}\Psi_2$ can be seen as the calibration associated to domain walls. The directions in which the D-brane is stretched go along for the ride so that the problem
reduces to one on $\mathbb{R}\times M$, where $\mathbb{R}$ represents the direction transverse to the domain wall, say $x^3$. Then, the tension of a BPS domain wall obtained by a D-brane
wrapping $(x^1,x^2)$ and an internal generalized cycle $(\Sigma,\calf)$ is given by\footnote{To simplify expressions, we will adopt units that put $2\pi\sqrt{\alpha^\prime}=1$.}
\bea
T_{\text{\tiny DW}}=2\pi\left| \int_\Sigma e^{3A-\Phi}\Psi_2|_\Sigma\wedge e^\calf\right|=
2\pi\left|\int_{\mathbb{R}\times M}\langle e^{3A-\Phi}\Psi_2, j^{\text{\tiny DW}}_{(\Sigma,\calf)}\rangle\right| \, ,
\eea
where $j^{\text{\tiny DW}}_{(\Sigma,\calf)}$ is the generalized current \cite{paulluca} in $\mathbb{R}\times M$  associated to the domain wall. Using the Bianchi identities,  $\d_H F=-j^{\text{\tiny DW}}_{(\Sigma,\calf)}$, we obtain
\bea\label{branedw}
T_{\text{\tiny DW}}=2\pi\left| \int_{\{x^3=\infty\}\times M}\langle e^{3A-\Phi}\Psi_2, F\rangle-\int_{\{x^3=-\infty\}\times M}\langle e^{3A-\Phi}\Psi_2, F\rangle\right|\ .
\eea
From the four-dimensional point of view, the D-brane can be seen as the BPS domain wall separating two flux configurations.  Thus the expression (\ref{branedw}) has to be compared with the four-dimensional formula (see \eqref{einsteindw} and surrounding explanation)
\bea\label{sugradw}
T_{\text{\tiny DW}}=2|\Delta\calw| \, ,
\eea
from which we get the following {\em on-shell} superpotential
\bea\label{sup1}
\calw_{\text{on-shell}}=\pi\int_{M}\langle e^{3A-\Phi}\Psi_2, F\rangle\ .
\eea

To extrapolate the above on-shell expression to the complete off-shell superpotential we demand holomorphicity as a minimal requirement.
We should thus first find the natural complex variables of the setup.

\subsection{Holomorphic variables and off-shell superpotential}

Let us start with the NSNS degrees of freedom. As recalled in section \ref{back}, the two pure spinors $\Psi_1$ and $\Psi_2$ define the internal metric $h$. In fact, as explained in appendix \ref{hodge} the $B$-twisted pure spinors, indicated with a prime, also contain information on the $B$-field, so that all the internal NSNS degrees of freedom  $h$, $B$ and $\Phi$ are contained in the two twisted pure spinors $e^{-\Phi}\Psi^\prime_1$ and $e^{-\Phi}\Psi^\prime_2$.

Next, the internal RR degrees of freedom are contained in the (locally defined) RR-potentials $C$, which satisfy $\d_H C=F$ (or their twisted counterpart $C^\prime$ such that $\d C^\prime=F^\prime$).\footnote{For massive IIA this expression is modified to $F=\d_H C + m e^{-B}$ where the mass $m=F_{(0)}$.
However, the integrand of the Chern-Simons part of the D-brane action $\theta_{\text{\tiny CS}}$ also changes in such a way that still $\d \theta_{\text{\tiny CS}} = F \wedge e^{\calf}$, so that the argument based on D-brane instantons we will make below is still valid.} Furthermore, let us keep in mind that,
as shown in \cite{hitchin}, the complex pure spinors are in fact completely defined by their real or imaginary part.

In our problem, a first natural holomorphic variable is suggested by the superpotential (\ref{sup1}) itself and is given by
\bea\label{hol1}
\Z^\prime \equiv e^{3A-\Phi}\Psi_2^\prime\ .
\eea

On the other hand, we see in (\ref{sup1}) the appearance of the RR field-strength $F$, which does not appear in a complex combination. A natural completion is suggested by the coupling of BPS D-brane instantons to the background fields. The proper supersymmetry conditions for D-brane instantons are studied in appendix \ref{appinst}.  The outcome is that the action of a supersymmetric D-brane instanton wrapping a supersymmetric generalized cycle $(\Sigma,\calf)$ in $M$ is
\bea
S_{\text{E}}=2\pi\int_{\Sigma}(e^{-\Phi}\Re \Psi_1-i C)|_\Sigma\wedge e^\calf\ .
\eea
Thus we see that the natural complexification of the RR field-strength $F$ is
\bea
F+i\,\d_H(e^{-\Phi}\Re \Psi_1)\ ,
\eea
and indeed, in the probe approximation we are using, the additional term does not modify the domain wall tension \eqref{branedw}. The same complex completion is obtained by considering space-filling branes and looking at the holomorphic function defining the vector multiplet kinetic term, see appendix \ref{appinst}.

To introduce the proper holomorphic variable we change to the twisted picture, fix a certain representative RR field-strength $F_0^\prime$ so that $F^\prime=F_0^\prime+\d\Delta C^\prime$, where $\Delta C^\prime$ is now a globally defined twisted polyform. Then one defines
\bea\label{hol2}
\T^\prime \equiv e^{-\Phi}\Re\Psi_1^\prime-i\Delta C^\prime\ .
\eea
That the holomorphic variable should take this form was already proposed in \cite{grimm} based on earlier work \cite{grimmlouis}.

For the following discussion,  it is useful to introduce also the pure spinor
\bea\label{t}
\t^\prime \equiv e^{-\Phi}\Psi_1^\prime\ ,
\eea
whose degrees of freedom can be considered as contained in $\T^\prime$, since  $\Im \t^\prime$ can be seen as a function of $\Re \t^\prime=\Re\T^\prime$. Note however that the complex structure induced by $\t^\prime$ on the degrees of freedoms contained in $\Re \t^\prime$ is by definition incompatible with the  complex structure induced by $\T^\prime$. In other words, the embedding  $\t^\prime\hookrightarrow \T^\prime$ is not holomorphic.

In terms of the new variables the constraint (\ref{psconst2}) can be written as
\bea\label{comp3}
\langle \mathbb{X}\cdot\Z^\prime,\Re\T^\prime\rangle =0 \, , \qquad \forall \mathbb{X}\in T_M\oplus T^\star_M\ ,
\eea
so that together $\Z^\prime$ and $\Re\T^\prime=\Re t^\prime$ provide an $SU(3)\times SU(3)$-structure -- and thus a metric $h$ and $B$-field -- for the twisted extension bundle (\ref{ext}). Moreover they also define the dilaton $\Phi$ via
\bea\label{dilaton}
e^{2\Phi}=\frac{4\sqrt{\det h}\,\d^6 y}{\langle \Re\t^\prime,\Im\t^\prime\rangle}\ ,
\eea
where again $\Im\t^\prime$ should be considered as a function of $\Re\t^\prime=\Re\T^\prime$, and the warp factor $A$ through
\bea\label{ccc}
e^{6A}=\frac{\langle \Z^\prime,\bar\Z^\prime\rangle}{\langle \t^\prime,\bar\t^\prime\rangle}\ .
\eea

Thus, we see that $\Z^\prime$ and $\T^\prime$ contain all the geometrical information about the compactification and can be thus considered as the chiral fields of the four-dimensional description. From (\ref{sup1}), by demanding holomorphicity with respect to $\Z^\prime$ and $\T^\prime$,  we finally arrive at the following manifestly holomorphic {\em off-shell} superpotential
\eq{
\label{classup0}
\calw = \pi\int_M\langle \Z^\prime,F_0^\prime+i\,\d\T^\prime\rangle\ .
}

In fact, their is some redundancy that we will explain later and plays a crucial role in the following discussions. Furthermore, it is important to remember that these variables are not completely independent but are constrained by (\ref{comp3}). The appropriate holomorphic parametrization of the fluctuations preserving this constraint is discussed in appendix \ref{infdef}.

From the four-dimensional point of view the twisted holomorphic variables defined above are conceptually the more appropriate.
However, from the ten-dimensional point of view the untwisted description is completely equivalent, merely transferring
the degrees of freedom of $B$ from inside the pure spinors into an explicitly appearing $H$-field. The untwisted picture
has the advantage that the pure spinors are globally defined and in the following we will prefer it, also to make easy contact
with the previous results reviewed in section \ref{back}. Anyway, the expressions are essentially identical,
as for going to the twisted picture the reader needs only to put primes on the polyforms and replace $\d_H$ by $\d$.

Thus, let us rewrite the off-shell superpotential (\ref{classup0}) in the untwisted picture
 \eq{
\label{classup}
\calw = \pi\int_M\langle \Z,F_0+i\,\d_H\T\rangle
      = \pi\int_M\langle \Z,F+i\,\d_H(\Re \T)\rangle\ .
}
This superpotential agrees with the ones obtained in the literature \cite{granaN2,grimm,granaN2bis} upon taking constant warp factor. Thus, as discussed in those papers (see also \cite{granareview}), it reproduces the different superpotentials found in the literature for particular subcases. We will come back to this point in section \ref{oldsup}.

\subsection{Intermezzo on open string degrees of freedom}

In this subsection we discuss how the open string superpotential of \cite{luca2} fits in. We note first that in the presence of back-reacting localized sources
-- space-filling D-branes or O-planes -- the RR flux must obey the Bianchi identities
\bea\label{sourcebianchi}
\d_H F=-j\ ,
\eea
where $j$ is the sum of the generalized currents \cite{paulluca,dimi} of all the sources: $j = \sum_{\text{D}p} j_{(\Sigma_p,\calf)} + \sum_{\text{O}p} j_{(\Sigma_p)}$.
Thus, the tadpole condition demands that $j$ must be $\d_H$-exact, which means that there must be a generalized chain \cite{lucajarah} whose generalized boundary gives the localized sources, and we indicate the corresponding current with $\theta$, such that $\d_H\theta=-j$. Then, we can split
\bea\label{rrsplit}
F=\theta+\hat F\ ,
\eea
so that $\d_H \hat F=0$. In this way we have isolated the open string degrees of freedom and moved them in $\theta$ so that $\hat F$ contains
only closed string degrees of freedom. Then, in the presence of D-branes an argument based on domain walls similar to that in subsection \ref{onshell} exists for the open string superpotential \cite{luca2}. Putting both together, the complete closed plus open string superpotential can be obtained by considering a generalized chain as defined in \cite{lucajarah} interpolating between two vacua with possibly different numbers of space-filling D-branes and background fluxes. The result is that the total superpotential has the form
\bea
\calw=\calw_{(c)} + \calw_{(o)}\ ,
\eea
where the closed string superpotential $\calw_{(c)}$ is like (\ref{classup}) with $\hat F$ instead of $F$, while in the case of isolated D-branes
\bea\label{osup}
\calw_{(o)}=\pi\int_M\langle\Z, \theta\rangle
\eea
exactly reproduces the D-brane superpotential found in \cite{luca2}.

Note however that, differently from the closed string superpotential, the open string superpotential (\ref{osup}) is only well-defined when the background has a flat four-dimensional part and is on-shell, so that $\d_H\Z=0$. We suspect that this problem comes from the fact that the $\kappa$-symmetric D-brane superaction \cite{ceder} is only well-defined when the background is on-shell. We therefore expect that a consistent introduction of D-brane degrees of freedom is easier once the closed string spectrum has been consistently truncated to the light modes. As an additional complication D-branes should also modify the K\"ahler potential. We will not address these problems here and explicitly consider space-filling D-branes only in section \ref{corrections}.

\subsection{Four-dimensional supergravity}

Let us start by observing that the metric ansatz (\ref{metric}) has an intrinsic ambiguity. Namely, we can simultaneously shift the warp factor and rescale the four-dimensional metric as follows
\bea\label{weyl}
A\rightarrow A+\sigma \quad\text{and} \quad  g \rightarrow e^{-2\sigma}g \ ,
\eea
 while preserving the ansatz (\ref{metric}). In particular, when we consider a generic configuration -- not necessarily a vacuum -- $\sigma$ can be an arbitrary function of the $x^\mu$ coordinates.

Now, the warp factor is given essentially by $\calz$ through (\ref{ccc}). In fact $\calz$ contains a redundancy associated to its overall phase. In particular we can perform the chiral rotation
\bea\label{chiral}
\calz\rightarrow e^{i\alpha}\calz\ ,
\eea
where $\alpha$ is in general $x^\mu$-dependent, without changing the geometrical content of the ten-dimensional configuration.

Putting together (\ref{weyl}) and (\ref{chiral}) we see that a four-dimensional description of this class of configurations in terms of $\calz$ and $\T$ should be naturally invariant under the local complexified Weyl transformations
\bea\label{complweyl}
\calz\rightarrow \Lambda^3\calz \quad\text{and} \quad  g \rightarrow |\Lambda|^{-2} g \ ,
\eea
where $\Lambda=e^{\sigma+\frac{i}{3}\alpha}$ is an arbitrary nowhere vanishing complex function of the $x^\mu$s,  and $\T$ remains invariant.

The general $N=1$ supergravity with such a gauge invariance can be constructed by partially gauge-fixing the superconformal action of \cite{supconf}\footnote{We thank Toine Van Proeyen for explaining this to us.}. We summarize here the main ingredients needed for our purpose, leaving some more details for appendix \ref{sugrapp}.

The four-dimensional supergravity action contains an Einstein term of the form
\bea\label{eins}
S=\frac{1}{2}\int_X \d^4x\sqrt{-\det g}\,\caln\, R + \ldots \, ,
\eea
where $R$ is the scalar curvature of the four-dimensional metric $g$ and $\caln$ can depend (non-holomorphically) on $\Z$ and $\T$. From dimensional reduction with ansatz (\ref{metric}) one can readily identify
\bea\label{warpedvol}
\caln=4\pi\int_M\d^6y\sqrt{\det h}\, e^{2A-2\Phi} \, .
\eea
This can be written in terms of $\Z$  and $\t$ (and thus $\T$) in the following equivalent ways
\bea\label{comp}
\caln&=&\frac{\pi i}{2}\int_{M}e^{-4A}\langle\Z,\bar\Z\rangle=\frac{\pi i}{2}\int_{M}e^{2A}\langle\t,\bar\t\rangle=\frac{\pi i}{2}\int_{M}\langle\t,\bar\t\rangle^{2/3}\langle\Z,\bar\Z\rangle^{1/3}=\cr
&=&\frac{\pi}{2}\Big(i\int_{M}e^{-4A}\langle\Z,\bar\Z\rangle\Big)^{1/3}\Big(i\int_{M}e^{2A}\langle\t,\bar\t\rangle\Big)^{2/3}\ .
\eea
where, in the first line, the rational powers of the two Mukai pairings combine to get a well-defined density. The last expression is useful for
taking first derivatives of $\caln$, since it turns out that upon using (\ref{ccc}) the contributions of
both factors conspire such that $A$ can be effectively considered as a constant. As we will see $\caln$ is related the
the K\"ahler potential in the usual Einstein-frame formalism, so we will call it the conformal K\"ahler potential.

The gravitino supersymmetry transformation contains the term
\bea\label{gravitino}
\delta\psi_\mu=\nabla_\mu\zeta_-+\frac{\calw}{2\caln}\,\gamma_\mu \zeta_++\ldots \, ,
\eea
from which, comparing with (\ref{spinori}),  we obtain the relation\footnote{The apparent sign discrepancy for type IIA can be fixed by changing the sign of the superpotential.}
\eq{
\label{w0}
W_0=\langle \calw/\caln\rangle \, .
}
Furthermore, the potential evaluated at a supersymmetric vacuum is given by
\bea\label{vpot}
\calv=-\frac{3|\calw|^2}{\caln}\ .
\eea
It follows that the cosmological constant is given by $\Lambda=\calv/\caln=-3|\calw|^2/\caln^2=-3|W_0|^2$.

The D-term $\cald(k)$  associated to a gauged Killing vector $k$ in the configuration space of the chiral fields, is given by
\bea\label{dterm}
\cald(k)= 3i k^{\text{hol}}(\caln)\ ,\eea
where $k^{\text{hol}}$ refers to the holomorphic projection of $k$.
As for the F-flatness equation associated to a chiral field $\phi$ we can find it from (\ref{finsusy}) and it has the form
\bea\label{fterm}
\partial_\phi\calw-3(\partial_\phi\log\caln)\calw=0\ .
\eea

Even if keeping the formulation explicitly invariant under the complexified local Weyl transformations (\ref{complweyl}) is more natural and allows to easily relate the four-dimensional expressions to the ten-dimensional ones, one can fix this gauge invariance as explained in \cite{supconf}. This leads to a standard $N=1$ supergravity in the Einstein frame with
\bea\label{pmass}
\caln=M_{\text{P}}^2\ ,
\eea
where $M_{\text{P}}$ is the four-dimensional Planck length measured in units where $2\pi\sqrt{\alpha^\prime}=1$.\footnote{Here and in the following formulae, to reintroduce the explicit dependence on the string length one must multiply  $M_P$ by $2\pi\sqrt{\alpha^\prime}$ in order to get a dimensionless quantity.} As in \cite{supconf},  one can explicitly isolate the `compensator' $Y$ in $\Z$ by redefining
\bea\label{splitting}
\Z\rightarrow Y\Z(z)\ ,
\eea
where now $\Z$ should be thought of as a section of a complex line bundle, whose holomorphic base coordinates we indicate with $z$. Indeed, there is an ambiguity in the splitting (\ref{splitting}) under the redefinition
\bea\label{red}
Y\rightarrow e^{-f(z)}Y\quad \text{and}\quad \Z\rightarrow e^{f(z)}\Z\ ,
\eea
with $f(z)$ an arbitrary holomorphic function.

In the new split variables the four-dimensional part of the ten-dimensional metric (\ref{metric}) becomes explicitly dependent on the compensator
\bea\label{metric2}
\d s^2=e^{2A}|Y|^2g+h\ ,
\eea
where $A$ depends on the new $\Z$ as in the old expression (\ref{ccc}).
Note that the transformation (\ref{red}) is not a complexified Weyl transformation since it does not affect the four-dimensional metric $g$, but only the warp factor and is balanced by the explicit appearance of the compensator $Y$ in (\ref{metric2}).

The superpotential $\calw_{\text{\tiny E}}$ of the resulting standard Einstein supergravity is
\bea\label{einsup}
\calw_{\text{\tiny E}}=M_{\text{P}}^3\,\calw \, ,
\eea
and the K\"ahler potential is given by
\bea\label{classk0}
\calk=-3\log\caln \, ,
\eea
where now $\calw$ and $\caln$, while still formally given by the old expressions (\ref{classup}) and (\ref{comp}), must be considered as functions of the new $\Z$ after the splitting (\ref{splitting}). From the Weyl gauge-fixing condition (\ref{pmass}) we get
\bea
|Y|^2=M_{\text{P}}^2\, e^{\calk/3}\ .
\eea
The invariance under the redefinition (\ref{red}) gives rise to the  K\"ahler invariance of the usual Einstein supergravity
\bea\label{kahlertr}
\Z\rightarrow e^f\Z\, ,\qquad \calk\rightarrow \calk -f-\bar f \quad\text{and} \quad \calw_{\text{\tiny E}}\rightarrow e^f \, \calw_{\text{\tiny E}}\ .
\eea

The chiral symmetry (\ref{chiral}) can be fixed by imposing $Y=\bar Y$ \cite{supconf}. Although this  breaks the  invariance of the theory under (\ref{red}), the K\"ahler symmetry can be preserved by considering it as a combination of (\ref{red}) and a correcting chiral transformation (\ref{chiral}).

From \eqref{comp} and (\ref{classk0}) we find that the K\"ahler potential can be written as
\bea\label{classk}
\calk&=&-2\log \Big(i\int_M e^{2A}\langle \t,\bar\t \rangle\Big)-\log \Big(i\int_M e^{-4A}\langle \Z,\bar{\Z} \rangle\Big) -3\log\frac{\pi}{2}\  \ .
\eea
However from (\ref{comp}) we also find that the K\"ahler potential may be expressed in alternative ways using the dependence of $A$ on $\Z$ and $\t$, which also shows explicitly how the non-trivial warp factor couples $\Z$ with $\t$ (and thus $\T$). This breaks a possible underlying $N=2$ description of the system where $\Z$ and $\t$ would belong to vector- an hypermultiplets respectively \cite{granaN2,granaN2bis} and decouple.

The form (\ref{classk}) for the K\"ahler potential is the natural one for considering the constant warp factor approximation, since in this form the warp factor effectively
disappears. This avoids problems related to the hidden dependence of the warp factor on $\Z$ and $\t$  and is consistent with K\"ahler covariance (\ref{kahlertr})
without worrying about the warp factor transformation. By putting $A$ constant in (\ref{classk}) and restricting to $SU(3)$-structure backgrounds, one gets indeed the K\"ahler potential obtained in \cite{grimm} by dimensional reduction. Furthermore, in this approximation one recognizes the underlying $N=2$ structure described in \cite{granaN2,granaN2bis}.

As a check, for constant warp factor, the ten-dimensional metric takes the form
\bea
\d s^2=\frac{M_{\text{P}}^2}{4\pi\text{Vol}(M)}g+h
\eea
where $\text{Vol}(M)$ is the {\em unwarped} volume of the internal manifold $M$, thus reproducing the standard relation
\bea
g^{\text{st}}=\frac{M_{\text{P}}^2}{4\pi\text{Vol}(M)}g\ ,
\eea
between the four-dimensional string- and Einstein-frame metrics for unwarped compactifications.

Note that with (\ref{pmass}), (\ref{einsup}) and (\ref{classk0}) the standard four-dimensional Einstein-frame formula
for the domain wall tension, which under suitable assumptions of the phases reads \cite{cvetic}
\bea
\label{einsteindw}
T_{\text{DW}}=2|\Delta(e^{\calk/2}\calw_{\text{\tiny E}})| \, ,
\eea
can indeed be rewritten as (\ref{sugradw}) in terms of the superpotential $\calw$ of the Weyl invariant description.

As a final comment, the K\"ahler potential (\ref{classk}) satisfies a no-scale type condition \cite{kounnas}
\bea\label{nonscale}
\frac{\partial\calk}{\partial \phi^I}(\calk^{-1})^{I}{}_J\frac{\partial\calk}{\partial \bar\phi_J}=4\ ,
\eea
where $\calk_I{}^J=\partial^2\calk/(\partial\phi^I\partial\bar\phi_J)$ and the $\phi^I$ represent a set of chiral fields obtained by expanding $\T$ in some base. The condition (\ref{nonscale}) is usually presented for constant warp factor, but here we see that it is valid in general. To derive it, we first rewrite the K\"ahler potential (\ref{classk}) as
\bea\label{classk2}
\calk=-3\log\Big(\frac{\pi i}{2}\int_M\langle\t,\bar t\rangle^{2/3}\langle\Z,\bar\Z\rangle^{1/3}\Big)\ .
\eea
 Then, the condition (\ref{nonscale}) follows from some simple algebraic manipulations, using the fact that the integrand in (\ref{classk2}) is homogeneous of degree $4/3$ under the rescaling $\Re\T=\Re\t\rightarrow \alpha \, \Re\T$.

\section{Equivalence of the supersymmetry conditions in the four- and ten-dimensional description}
\label{10susy}

In this section we discuss how the four-dimensional vacuum supersymmetry conditions obtained from the conformal K\"ahler potential $\caln$ and
the superpotential $\calw$ reproduce the full ten-dimensional supersymmetry conditions reviewed in section \ref{back}. We first discuss D-terms and the resulting D-flatness condition in subsection \ref{dterms}. The F-flatness conditions are discussed
next in subsection \ref{adscomp}. In the AdS case, they automatically imply the D-flatness condition, in agreement with standard four-dimensional arguments.

\subsection{D-terms in generalized flux compactifications}
\label{dterms}

There is some redundancy in the parametrization of the degrees of freedom with $\T$ and $\Z$, which is given by the RR gauge transformations $\delta_\lambda C=\d_H\lambda$. They induce the transformation $\delta_{\lambda}\T=-i\d_H\lambda$.  These transformations can be seen as symmetries of the field space, which are gauged in the final theory.

Indeed, consider the vector-like RR-fields $C_{[1]}$ with one leg along $X$.  From the four-dimensional point of view they correspond to gauge fields and the associated   gauge transformations are obtained by `gauging' $\lambda$ to depend also on the coordinates $x^\mu$ of $X$.  Then the gauge transformations are given by $\delta_{\lambda}C_{[1]}=\d_x\lambda$ and $\delta_{\lambda}C=\d_H\lambda$, where $\d_x=\d x^\mu\partial_{\mu}$ and $\d_H=\d y^m\partial_{m}+H \wedge$ are the exterior derivatives along $X$ and $M$ respectively. Of course, these axionic gauge symmetries would be spontaneously broken in any vacuum, so that these fields gain a mass by the Higgs mechanism.

On general grounds, we expect the existence of D-terms corresponding to these gauged symmetries.\footnote{Note that these gauge symmetries are invisible in a reduced four-dimensional theory, which uses only cohomology representatives.  In such a reduced theory the massive modes are supposed to be integrated out and only zero modes are considered.}${}^{,}$\footnote{For IIA $SU(3)$-structure toroidal vacua a previous analysis of D-terms
from the ten-dimensional point of view can be found in \cite{robbins}.}  From $\delta_\lambda\T=-i\d_H\lambda$, we can extract the form of the D-term using (\ref{dterm}), obtaining
\bea
\cald(\lambda)=2\pi\int_M\langle \lambda, \cald\rangle \, ,
\eea
with the D-term density $\cald$ given by
\bea\label{dtermt}
\cald=\d_H(e^{2A}\Im \t)\ .
\eea

From (\ref{t}) we see that the supersymmetry equation (\ref{intcond1}) can be interpreted as a D-flatness condition
\bea\label{dflat}
\cald=0\ .
\eea

\subsection{F-terms and ten-dimensional supersymmetry}
\label{adscomp}

In this subsection we show that the ten-dimensional supersymmetry equations \eqref{susyconds} and \eqref{intconds} can be obtained from the superpotential (\ref{classup}) and the conformal K\"ahler potential (\ref{comp}).
Firstly, we parameterize the independent fluctuations of our fields $\T$ and $\Z$ as discussed in appendix \ref{infdef}, taking into account that $\Re\T$ and $\Z$ are subject to the constraint  (\ref{comp3}). This means that they parameterize the possible $SU(3)\times SU(3)$-structures on $M$ and are not completely independent, while $C=-\Im\T$ on the other hand is independent.
The  result is that the independent holomorphic deformations of $\T$ and $\Z$ are classified
by the Hodge decomposition induced by the (almost) generalized complex structure associated to $\calz$ as follows
\bea\label{varfields}
\delta\T\in\Gamma(V_{0}\oplus V_{-2})\quad\text{and}\quad \delta\Z\in\Gamma(V_{3}\oplus V_{1})\ .
\eea

Let us consider first the variation $\delta\T$, for which the F-flatness condition
\bea\label{FT}
\delta_{\T}\calw-3(\delta_{\T}\log\caln)\,\calw=0
\eea
can be readily seen to give rise to the equation
\bea\label{altcond}
\d_H\Z =\frac{2i\calw}{\caln}\, e^{2A} \, \Im\t=2i\,W_0\,e^{2A} \, \Im\t\ ,
\eea
where we have used \eqref{w0}. This exactly reproduces the complex combination of (\ref{susycond2}) and \eqref{intcond2}. In the generic AdS case, since $W_0\neq 0$ this equation implies (\ref{intcond1}), which we have already interpreted as a D-flatness condition in the previous subsection.\footnote{Note that it also implies (\ref{comp3}), using the fact that in general $\langle \mathbb{X}\cdot \Z,\d_H\Z\rangle\equiv 0$.}

Secondly, the F-flatness equation resulting from a variation of $\Z$ is
\bea\label{var1}
\int_M\langle \delta\Z,\d_H\Re\t -iF-\frac{\calw}{2\caln} \, e^{-4A}\bar\Z \rangle=0\ .
\eea
Considering the two different deformations in (\ref{varfields}) separately, and using again  $W_0=\langle \calw/\caln\rangle$ one gets\footnote{In the
case $W_0=0$ these equations were already rewritten in this form in \cite{alessandro}.}
\begin{subequations}
\label{susy2}
\begin{align}
\label{susy21}
&(\d_H\Re\t -iF)|_{V_{-1}}=0\, , \\
\label{susy22}
&(\d_H\Re\t -iF)|_{V_{-3}}=\frac12 W_0 \, e^{-4A} \bar\Z\ .
\end{align}
\end{subequations}
In order to show the equivalence between (\ref{susy2}) and (\ref{susycond1}) one has to use the double generalized Hodge decomposition of appendix \ref{hodge}
and the associated Hodge duality property \eqref{hodgedual}. The $U_{0,-1}$-component of \eqref{susy21} is then readily seen to be equivalent to the corresponding
component in \eqref{susycond1} as are the $U_{\pm 2,-1}$-components upon using
\eq{
\d_H (e^{2A} \Re t)|_{U_{\pm 2,-1}} =0 \, ,
}
which follows from \eqref{intcond1} using \eqref{nonintdec}. Finally, to show the equivalence of the $V_{-3}$-component one
first notes that (\ref{susy22}) implies, together with \eqref{altcond}, that
\bea
\label{F3}
F|_{V_{-3}}=-\frac{i}{2}\, W_0\, e^{-4A} \bar\Z\ .
\eea
We can thus rewrite it as
 \bea
 (\d_H\Re\t +iF)|_{V_{-3}}=\frac{3}{2} \, W_0\,e^{-4A} \bar\Z\ ,
 \eea
which we find, using again the Hodge duality property, to be equivalent
to the corresponding component of (\ref{susycond1}).

This closes our proof that for AdS compactifications the F-flatness conditions arising from the superpotential (\ref{classup}) and conformal K\"ahler potential (\ref{comp}) reproduce the complete set of ten-dimensional supersymmetry equations (\ref{susyconds}) and (\ref{intconds}). In the case of compactifications to flat $\mathbb{R}^{1,3}$ one has to add the D-flatness condition (\ref{dflat}) which is then not automatically implied by the F-flatness condition (\ref{altcond}).

As a final remark, let us note that if we would have tried to restrict from the beginning to generalized complex vacua (such that $\d_H\calz=0$) the superpotential (\ref{classup}) would have been completely independent of $\T$. Thus, due to the no-scale property (\ref{nonscale}) the effective potential would have been positive definite, and vanishing at the supersymmetric vacuum \cite{kounnas}, leading to a vanishing cosmological constant, in agreement with our previous ten-dimensional considerations. This is very similar to what happens in the usual approach to warped Calabi-Yau  compactifications, which we will revisit in the following section.

\section{Revisiting the Gukov-Vafa-Witten superpotential}
\label{oldsup}

In the previous section we have proven that the four-dimensional superpotential (\ref{classup}) and the conformal K\"ahler potential (\ref{comp}) (or the K\"ahler potential (\ref{classk}) in the gauge-fixed Einstein-frame description) reproduce the full ten-dimensional supersymmetry equations. A key point in the derivation of these D- and F-flatness conditions is that in fact we consider the full Kaluza-Klein spectrum of fluctuation modes. This is in contrast with the path usually followed when studying the moduli-lifting upon adding fluxes to a Calabi-Yau manifold. The warp factor is then often approximated by a constant and the spectrum of internal fluctuations is truncated to what would be the massless spectrum of the underling Calabi-Yau geometry. The superpotential is then seen as a function of only these `light' modes.

Prototypical for this approach is the famous Gukov-Vafa-Witten superpotential for IIB warped Calabi-Yau compactifications
\bea\label{gvw}
\calw_{\text{GVW}}=\int_M\hat\Omega\wedge G_{(3)}\ ,
\eea
where $\hat\Omega$ is the holomorphic $(3,0)$-form associated to the Calabi-Yau geometry and  $G_{(3)}=F_{(3)}+ie^{-\Phi}H=\d C_{(2)}+\tau H$, with $\tau=C_{(0)}+ie^{-\Phi}$ constant. On the other hand the K\"ahler potential is obtained essentially from the underlying $N=2$ theory, by truncating the spectrum via the O3/O7 orientifold projection and neglecting the warp factor. For the K\"ahler moduli one has to identify the right holomorphic variables as in \cite{grimmlouis}. This gives a no-scale theory, with corresponding supersymmetry  equations
\bea
G^{3,0}=G^{0,3}=G^{1,2}=0\ .
\eea
 This however does not reproduce all the ten-dimensional supersymmetry equations. Indeed, one has to add two additional conditions. The first is the primitivity condition
\bea
\hat J\wedge G=0\ ,
\eea
where $\hat J$ denotes the  K\"ahler structure of the Calabi-Yau. The second is a relation between $F_{(5)}$ and the warp factor, which we will write out later. Furthermore, one can extend this class of warped Calabi-Yau compactifications to F-theory compactifications, with non-constant holomorphic axion-dilaton $\tau$. This case also seems not to be covered by (\ref{gvw}).

To clarify the general but somewhat formal analysis of the previous sections, let us repeat it in detail for the more familiar example of warped Calabi-Yau compactifications. They can be seen as a particular subclass of the (strict) $SU(3)$-structure compactifications, which are in turn obtained as a subsector of the more general $SU(3)\times SU(3)$ vacua considered in this paper by setting
\bea\label{su3ansatz}
\Psi^{+}=e^{i\vartheta} e^{iJ}\quad\text{and}\quad \Psi^-=\Omega\ ,
\eea
where $\vartheta$ is in general point-dependent and $J$ and $\Omega$ define the $SU(3)$-structure, for which the conditions (\ref{psconsts}) become
\bea
J\wedge \Omega=0 \, , \qquad \frac{1}{3!}J\wedge J\wedge J=-\frac{i}{8}\Omega\wedge\bar\Omega=\sqrt{\det h}\,\d^6y\ .
\eea

For the IIB warped Calabi-Yau compactifications $e^{i\vartheta}=1$ so that the `normalized' pure spinors take the form
\bea\label{cyps}
\Z\equiv\hat\Omega= e^{3A-\Phi}\Omega\quad\text{and}\quad \t=e^{-\Phi}e^{iJ}\ ,
\eea
and thus
\bea\label{cybigt}
\T=e^{-\Phi}\Re(e^{iJ})-i\Delta C\ .
\eea

By plugging (\ref{cyps}) and (\ref{cybigt}) into  (\ref{classup}) one indeed gets (\ref{gvw}).  Furthermore, the K\"ahler potential (\ref{classk}) takes the form
\bea
\calk_{\text{wCY}}=-2\log \Big(\frac43 \int_M e^{2A-2\Phi}J\wedge J\wedge J\Big)-\log\Big(-i\int_M e^{-4A}\hat\Omega\wedge\overline{\hat\Omega}\,\Big) -3\log\frac{\pi}{2}\ ,
\eea
which, for constant dilaton and under the identification $\hat J=e^{2A-\Phi} J$, agrees with the warped K\"ahler potential proposed in \cite{giddings1,giddings2}.

From our general discussion, we know that the superpotential (\ref{classup}) and the K\"ahler potential  (\ref{classk}), as functionals of all the fluctuation modes, reproduce the full set of ten-dimensional supersymmetry conditions. Therefore, in our approach we start by  assuming only the $SU(3)$-structure condition (\ref{cyps}), with no integrability of the complex and (warped) K\"ahler structures. Consider first the D-flatness condition. As we have seen it is associated to the RR gauge transformations $\delta C=\d_H\lambda$ and is thus invisible in an approach considering only cohomology representatives. From (\ref{dtermt}) we have the equation
\bea
\d_H(e^{2A-\Phi}J)=0\ .
\eea
which implies that $\hat J=e^{2A-\Phi}J$ is closed (so that the internal space, if complex, has a warped K\"ahler metric) and $H$ is primitive. So we naturally obtain as a D-flatness condition what should be imposed by hand in the usual approach\footnote{However, in specific cases, it can still appear from an effective four-dimensional gauged supergravity approach, see e.g. \cite{marioD}.}.

Let us now turn to the F-flatness conditions, starting from those of the form (\ref{FT}). Since we want to consider solutions which respect the ansatz (\ref{cyps}), we must use
the generalized Hodge decomposition $\Lambda^\bullet T^\star_M\otimes \mathbb{C}=\bigoplus_k V_k$ associated to
$\Z=\hat\Omega$ to decompose the space of complex forms. Its relation to the standard Hodge decomposition is
\bea
V_k=\bigoplus_p\Lambda^{3-p,3-k-p}\ ,
\eea
where $\Lambda^{p,q}$ contains the $(p,q)$-forms as defined by the (for the moment almost-) complex structure $\hat\Omega$. Then, from the discussion in appendix \ref{infdef} a set of holomorphic independent deformations of $\T$ is given by
\bea
\delta\T&\in& V_{-2}=\Lambda^{1,3}\oplus \Lambda^{0,2}\ ,\cr
\delta\T&\in& V_0=\Lambda^{0,0}\oplus \Lambda^{1,1}\oplus \Lambda^{2,2}\oplus \Lambda^{3,3}\ .
\eea

Since $\Z=\hat\Omega$,  it is easy to see that $\delta\T^{1,3}$ and $\delta\T^{3,3}$  give empty F-flatness conditions, while $\delta\T^{1,1}$ and $\delta\T^{0,2}$ give the conditions
\bea
(\d\hat\Omega)^{2,2}=0\quad \text{and}\quad(\d\hat\Omega)^{3,1}=0\ .
\eea
The first implies that $\hat\Omega$ defines an integrable complex structure, and the second that $\hat\Omega$ is holomorphic with respect to it. These conditions are also usually assumed from the start while here they are derived from the superpotential (\ref{classk}).

Secondly $\delta\T^{2,2}$ gives the F-flatness condition
\bea\label{noscale}
(\calw/\caln) \, \hat J=0\ ,
\eea
which in turn implies that $W_0=\langle \calw/\caln\rangle=0$, so that it is impossible to have AdS vacua satisfying the ansatz (\ref{cyps}) and (\ref{cybigt}), and in fact the general IIB $SU(3)$-structure ansatz (\ref{su3ansatz}). This ten-dimensional result is usually associated to the no-scale structure \cite{kounnas} that is observed  in the effective theory with truncated spectrum and constant warp factor \cite{gkp}. From the discussion at the end of the previous section, this no-scale property remains valid when taking into account a non-trivial warp factor.\footnote{In the previous section we discussed how the  no-scale property is valid for any compactification on a  generalized complex manifold such that $\d_H\calz=0$. In the warped Calabi-Yau case, one usually imposes the weaker condition  $\d\hat\Omega=0$, i.e.\ without also $H \wedge \hat\Omega=0$, so that the resulting superpotential cannot be considered independent of the axion-dilaton $\tau$. The only effect is that one needs to exclude the contribution of $\tau$ from the sum in the left-hand side of (\ref{nonscale}), leaving a $3$ on the right-hand side and thus still giving a no-scale theory.}

Finally we consider $\delta\T^{0,0}=-i\delta\tau$. The resulting equation is
\bea
\Omega\wedge H=0\quad\Rightarrow\quad H^{3,0}=H^{0,3}=0\ .
\eea
This condition can also be obtained (in cohomology) in the truncated theory by varying the Gukov-Vafa-Witten superpotential with respect to $\tau$.

Let us now turn to the F-flatness equations associated to $\Z$, see eq.~(\ref{var1}). From $\delta\Z^{3,0}$ and $\delta\Z^{2,1}$
one gets respectively
\bea
G^{0,3}=0\quad\text{and}\quad G^{1,2}=0\ .
\eea
These conditions can be also obtained in cohomology in the truncated theory as we reviewed before. However, in the untruncated theory we must consider two further kinds of fluctuations $\delta\Z^{3,2}$ and $\delta\Z^{1,0}$ that have no representatives in the truncated theory, since they correspond to `generalized complex' deformations. They give respectively
\bea\label{missing}
\bar\partial\tau=0\qquad\text{and}\qquad e^{\Phi} F^{2,3}=-\frac{i}{2}\bar{\partial}(4A-\Phi) \, J\wedge J\ .
\eea
The second condition can be re-expressed in terms of the `magnetic' dual of $F_{(5)}$ with four legs along $\mathbb{R}^{1,3}$, relating it to the warp factor and dilaton.  It is well-known that the two conditions (\ref{missing})  must be imposed in the full ten-dimensional solution, see e.g.\ \cite{granareview}, but are missing in the truncated approach. For an overview of all the supersymmetry equations of the warped Calabi-Yau setting, see table \ref{IIBsusy}.
\begin{table}
\begin{center}
\begin{tabular}{|c|c|c|}
\hline
Equation & Truncated theory (GVW) & Untruncated theory  \\ \hline
$\d \hat{\Omega}^{}=0$ \rule[1.1em]{0pt}{0pt} & assumed & F-flatness $\delta \T$ \\
$\d \hat{J}=0$ & assumed & D-flatness \\
$G^{0,3}=G^{1,2}=0$ & F-flatness & F-flatness $\delta \Z$ \\
$H^{3,0}=0$ & F-flatness & F-flatness $\delta \T$ \\
$H \wedge \hat{J}=0$ & added by hand & D-flatness \\
$\bar{\partial} \tau=0$ & added by hand & F-flatness $\delta \Z$ (generalized) \\
$4 \d A - \d \Phi = e^\Phi \star_6 \! F_{(5)}$ & added by hand & F-flatness $\delta \Z$ (generalized) \\ \hline
\end{tabular}
\end{center}
\caption{Derivation of the supersymmetry equations in IIB warped Calabi-Yau.\label{IIBsusy}}
\end{table}

Thus we see that, even if (\ref{classup}) reduces to (\ref{gvw}) once truncated, the truncated theory misses important information about the full ten-dimensional geometry,  which is encoded in (\ref{classup}) and (\ref{classk}) by considering the most general infinitesimal deformations including those in the `generalized complex directions'. Of course, the discussion presented in the previous sections is completely general and thus the analysis of this section can be repeated for (massive) IIA $SU(3)$-structure flux compactifications. In particular, in an $O6$-orientifold compactification we have the pure spinor ansatz
\bea
\Z=e^{3A-\Phi}e^{iJ}\, ,\qquad \t=e^{-\Phi}\Omega\ ,
\eea
and the truncated  superpotential (\ref{classup})  takes the form
\bea\label{IIa}
\calw_{\text{IIA}}=\int_M \big[\d(e^{3A-\Phi} J)-ie^{3A-\Phi}H\big]\wedge (e^{-\Phi}\Re\Omega) -\int_M F^\prime\wedge e^{3A-\Phi}e^{-(B+iJ)}\ ,
\eea
which for constant warp factor takes exactly the form found in \cite{gukov,micu1,zwirner1} for flux Calabi-Yau compactifications. Note however that here, as opposed to the IIB case considered above, the truncation of the theory based on Calabi-Yau geometry is not justified since one already knows that a generic\footnote{With a precise choice of orientifold sources, which have to be smeared however, it is possible to obtain
a Calabi-Yau solution \cite{acharya}.} supersymmetric IIA $SU(3)$-structure vacuum cannot have an underlying Calabi-Yau structure. It follows that one cannot `expand' the complete superpotential (\ref{classup}) around such a point. For IIB warped Calabi-Yau such an expansion, even if not complete, is at least consistent.

\section{Quantum corrected ten-dimensional geometry}
\label{corrections}

In sections \ref{dterms} and \ref{adscomp}  we have seen that the classical ten-dimensional supersymmetry equations (\ref{susyconds}) and (\ref{intconds})  can be reproduced as D- and F-flatness conditions of the four-dimensional effective theory. Note that each of these equations has a natural interpretation in terms of different
types of D-branes -- space-filling, domain-wall or string-like in four-dimensions -- `probing' the ten-dimensional space-time. Indeed, physical arguments require that a supersymmetric background be equipped with corresponding generalized calibrations of the kind introduced in \cite{gencalpaul,luca1}, fixing completely the structure of (\ref{susyconds}) and (\ref{intconds}). This result was shown for compactifications to flat space in \cite{luca1,luca2,lucajarah} and can be extended to AdS compactifications \cite{adsbranes}.  So we expect the main structure of (\ref{susyconds}) and (\ref{intconds}) to be preserved by both perturbative and non-perturbative quantum corrections, the latter being induced by world-sheet or brane instantons of different kinds. The four-dimensional point of view seems the most natural for studying such corrections. Thus, we start from there and use the formalism of the previous sections to understand the effect of the quantum corrections to the four-dimensional description in terms of corrections to the internal geometry of the full ten-dimensional picture.\footnote{The influence of non-perturbative corrections on
the internal structure of the compactification has already been considered in \cite{frey,lust}.}

The K\"ahler potential is expected to be affected by both perturbative and non-per\-tur\-ba\-tive corrections, see e.g.\ \cite{nonpert,grimmnonpert}. These are difficult to compute even in the simplest explicit settings, see e.g.~\cite{granareview,morereviews}, and we have not much to say about them in our general analysis. A possible guess is that they modify the (warped) Hitchin functionals entering the definition of the K\"ahler potential (\ref{classk}) and the constraint (\ref{ccc})  in such a way that from the F-terms and D-terms one can still write equations which have morally the same structure as (\ref{susyconds}) and (\ref{intconds}). In the following we will not consider such corrections explicitly and just use the classical K\"ahler potential (\ref{classk}).

We can say more about the corrections that may affect the superpotential. For the Gukov-Vafa-Witten superpotential (\ref{gvw}) at least, {\em perturbative} corrections are shown not to arise \cite{nonrenormGVW,vbb}. The argument of \cite{vbb} for instance is based on the non-renormalization of the tension of the BPS domain walls used in \cite{gukov} to derive this superpotential. Exactly this domain wall method we also used to extend the superpotential to the generalized case. While this might be thus one possible approach to a generalization of the non-renormalization theorem, this generalization does not seem straightforward to us, although this point clearly deserves a thorough study.

Well-known mechanisms that do generate {\em non-perturbative} corrections are Euclidean D-brane instantons (sometimes called E-branes) or stacks of space-filling D-branes that undergo gaugino condensation. Their supersymmetry conditions on general  $N=1$ flux vacua are studied in appendix \ref{appinst} and \cite{luca1,luca2} respectively, and it turns out that they must wrap the same internal generalized cycles.  Assume for the moment that there are no background branes different from the ones directly involved in the generation of the non-perturbative effect. Suppose these D-branes wrap an internal supersymmetric generalized cycle $(\Sigma,\calf)$, which we will refer to as the instantonic (generalized) cycle.
Then, using \eqref{instaction} and \eqref{gaugeF} we find for both mechanisms that the non-perturbative correction to the superpotential has the following form
\bea\label{npsup}
\calw_{\text{np}}=\cala\, \exp\Big( -\frac{2\pi}{n}\int_\Sigma \T|_{\Sigma}\wedge e^{\calf}\Big)\exp\Big( \frac{2\pi i}{n}\int_{\calb}F_0|_\calb\wedge e^{\tilde\calf}\Big)\ ,
\eea
where $n$ is the number of space-filling D-branes in the stack or $n=1$ for the instantonic D-brane contribution, and $(\calb,\tilde\calf)$ is a generalized chain whose generalized  boundary \cite{lucajarah} consists of $(\Sigma,\calf)$ and a fixed reference generalized cycle in the same generalized homology class \cite{lucajarah}. It is useful to introduce the generalized currents $j_{\text{np}}$ and $\theta_{\text{np}}$ associated to the cycle $(\Sigma,\calf)$  and the chain $(\calb,\tilde\calf)$ (so that $\d_H \theta_{\text{np}}=j_{\text{np}}-j_{\text{ref}}$). Then the non-perturbative correction to the superpotential (\ref{npsup}) can be written as
\bea\label{npsup2}
\calw_{\text{np}}=\cala\, \exp\Big( -\frac{2\pi}{n}\int_M \langle \T,\, j_{\text{np}}\rangle\Big)\exp\Big( \frac{2\pi i}{n}\int_M\langle F_0, \theta_{\text{np}}\rangle\Big)\ .
\eea

At one loop, the overall factor $\cala$ comes from the determinant of the Dirac action for the fermions on the D-branes \cite{fermions}, and should depend holomorphically on the background closed and open string degrees of freedom in a way consistent with the complexified Weyl invariance (\ref{complweyl}), or the K\"ahler invariance (\ref{kahlertr}) in the Einstein frame. Its explicit form is generically hard to compute, even if it may be possible to obtain a criterium for this factor not to vanish in terms of an index for the fermionic zero modes, extending the arithmetic genus considered for F-theory compactifications in \cite{witteninst}, similarly to the proposal presented in \cite{ale}. We will not attempt to address these important issues here and we will systematically assume that the dependence of $\cala$ on the closed string degrees of freedom is negligible for the following discussion. A possible dependence of $\cala$ on background space-filling D-branes will be considered in subsection \ref{instdef}. It turns out that even with this rather crude approximation we obtain nevertheless sensible results on the ten-dimensional geometry produced by the non-perturbative effect.\footnote{Let us also observe that the overall procedure to compute $\cala$ seems in principle not completely well-defined in our case, since it assumes the use of a probe instantonic D-brane on a fixed supersymmetric background vacuum and the validity of the extrapolation of the result to a general off-shell expression is not obvious.}

Now, repeating the analysis of section \ref{adscomp}, one easily finds that the non-perturbative contribution (\ref{npsup2}) generates a correction to eq.~(\ref{altcond})
\bea\label{gcscorr}
\d_H\Z=\frac{2i\calw_{\text{tot}}}{\caln}\, e^{2A}\Im\t+\frac{2i}{n}\,\calw_{\text{np}}\, j_{\text{np}}\ .
\eea
where we have used the total superpotential
\bea\label{totsup}
\calw_{\text{tot}}=\calw+\calw_{\text{np}}\ ,
\eea
while in (\ref{susy2}) one only needs to make the substitution $\calw\rightarrow \calw_{\text{tot}}$.

In the next subsections we will show that the modified equation (\ref{gcscorr}) provides insight in the ten-dimensional geometry of two important non-perturbative effects in IIB warped Calabi-Yau compactifications, which are usually only understood from the four-dimensional effective theory.  The first is the existence of supersymmetric AdS vacua resulting from the non-perturbatively corrected superpotential \cite{kklt}. The second is the
non-perturbative generation of a non-trivial superpotential for so-called mobile space-filling D3-branes, thereby reducing their moduli space.

\subsection{KKLT-like AdS vacua from smeared instantons}

Instead of focusing on the IIB warped Calabi-Yau case, let us be more general. Consider a classical supersymmetric compactification to flat $\mathbb{R}^{1,3}$. Thus $\calz$ is $\d_H$-closed
\bea
\d_H\calz=0\ ,
\eea
which is associated to the existence of an integrable generalized complex structure, reducing to the ordinary complex structure in the warped Calabi-Yau case.

Let us now include the non-perturbative effect, as explained above. Since $j_{\text{np}}$ is a localized source, clearly the perturbed equation (\ref{gcscorr}) can not be satisfied if $\d_H\calz=0$, implying that the generalized complex structure defined by $\calz$ cannot be integrable. However, the analysis of \cite{kklt} suggests a four-dimensional  mechanism to obtain an AdS vacuum in a warped Calabi-Yau compactification preserving its integrable complex structure. So the two points of view seem incompatible.

However one can consider a simplified version of \eqref{gcscorr} obtained by {\em smearing} the instanton generating the source term.
It is natural to introduce a supersymmetric smeared current as follows
\eq{
\tilde\jmath_{\text{np}} = \frac{\pi\sigma}{\caln}\,\, e^{2A}\Im t \ ,
}
where, in analogy with \cite{kklt}, we have defined
\bea
\sigma=\int_{\Sigma}\Re\T|_{\Sigma}\wedge e^\calf\ .
\eea
Indeed, $\tilde{\jmath}_{\text{np}}$ satisfies the calibration/supersymmetry conditions  $\tilde{\jmath}_{\text{np}} \in \Gamma(V_0)$ and $\langle \tilde{\jmath}_{\text{np}}, \Im \t \rangle=0$, see \cite{luca1,paulluca} and appendix \ref{appinst}. Furthermore the normalization is chosen such that
\bea\label{normcond}
\int_M\langle \Re\T,\tilde{\jmath}_{\text{np}}\rangle = \int_M\langle \Re\T,j_{\text{np}}\rangle =  \sigma\ .
\eea

Let us assume, like in \cite{kklt}, that $\calw$, $\cala$ and $\calw_{\text{np}}$ all have the same phase.  Then we see that substituting $\tilde\jmath_{\text{np}}$
for $j_{\text{np}}$ in (\ref{gcscorr}), we can now have a solution preserving the classical condition $\d_H\calz=0$. Indeed we just have to put the right hand side of (\ref{gcscorr}) to zero, finding
\bea\label{adsclass}
\calw=-\calw_{\text{np}}\big(1+ \frac{\pi}{n}\sigma\big)\ .
\eea
In (\ref{adsclass}) $\calw$ refers to the expectation value of the classical superpotential (\ref{classup}) that should be tuned to a small negative number by an appropriate choice of fluxes. This choice of fluxes should then be compatible with the appropriately quantum corrected equation (\ref{susy2}). That point is not under control yet. Anyway, eq.~(\ref{adsclass}) reproduces the structure of eq.~(13) of \cite{kklt}. In practice, the smearing puts us in the zero-mode approximation, implicit in the four-dimensional approach of \cite{kklt}.

Eq.~(13) of \cite{kklt} has an additional factor of $4/3$ multiplying $\sigma$ in (\ref{adsclass}). This discrepancy can be understood in the following way. In the case of a warped Calabi-Yau compactification as in \cite{kklt}, when we smear a four-dimensional generalized cycle $(\Sigma,\calf)$ as above, we assume the presence of a non-trivial world-volume flux $\calf$. This is generically the case since $H\neq 0$ and is needed in order for the smearing
\bea\label{warpedsmearing}
j_{\text{np}}&=&-\delta^{(2)}(\Sigma)+\delta^{(2)}(\Sigma)\wedge\calf-\frac12\delta^{(2)}(\Sigma)\wedge\calf\wedge\calf\cr &\rightarrow& \tilde\jmath_{\text{np}}=
\calc \, e^{2A-\Phi}(J-\frac1{3!}J\wedge J\wedge J)
\eea
to make sense. As above the normalization \eqref{normcond} puts then $\calc=\pi\sigma/\caln$.

However in \cite{kklt} it is assumed that $\calf=0$. This means that $j^{\text{\tiny KKLT}}_{\text{np}}=-\delta^{(2)}(\Sigma)$ and thus a more natural smearing would be
\bea\label{kkltsmear}
j^{\text{\tiny KKLT}}_{\text{np}}=-\delta^{(2)}(\Sigma)\quad\rightarrow \quad \tilde\jmath^{\,\text{\tiny KKLT}}_{\text{np}}=\calc_{\text{\tiny KKLT}}\, e^{2A-\Phi}J\ .
\eea
Imposing (\ref{normcond}) for $\tilde\jmath^{\,\text{\tiny KKLT}}_{\text{np}}$ determines $\calc_{\text{\tiny KKLT}}=4\pi\sigma/(3\caln)=4\calc/3$. This gives  exactly  the result of \cite{kklt}
\bea\label{kklt}
\calw_{\text{\tiny KKLT}}=-\calw_{\text{np}}\big(1+ \frac{4\pi}{3n}\sigma\big)\ .
\eea
Note however that one should in principle make the complete right-hand side of (\ref{gcscorr}) vanish, and the smearing (\ref{kkltsmear}) only accomplishes this partly. Thus,  our general ten-dimensional analysis suggests that a non-trivial world-volume flux $\calf$ on the instanton should be considered, leading to (\ref{adsclass}). An alternative to get (\ref{kklt}), keeping the smearing (\ref{kkltsmear}) and still satisfying (\ref{gcscorr}), may be possible by fine-tuning an appropriate $H$-field, necessarily such that $H\wedge \hat\Omega\neq 0$.

\subsection{Instanton generated generalized complex deformation and D3-brane moduli lifting}
\label{instdef}

Let us now consider a different application of (\ref{gcscorr}) based on localized instantons. In the regime of reliability of the non-perturbative correction, we expect the first term on the right-hand side of \eqref{gcscorr} to be small, as the estimate of the previous subsection confirms. Thus, locally, we may consider the source
term as the leading one on the right-hand side of  (\ref{gcscorr}) and write
\bea\label{gcscorr2}
\d_H\Z\simeq\frac{2i}{n}\,\calw_{\text{np}}\, j_{\text{np}}\ .
\eea
It follows that in this approximation the internal space can still be considered generalized complex away from the source, but not globally due to
the non-perturbative correction generating a localized obstruction term!

One interesting consequence of this remarkable result is obtained by considering again the warped Calabi-Yau IIB background. Then $\Z\equiv\hat\Omega$ is a holomorphic $(3,0)$-form defining an ordinary complex structure on $M$. In this background probe D3-branes have trivial classical superpotential $\calw_{\text{D3}}$, since $\d \calw_{\text{D3}}=-\pi \Z_{(1)}=0$ \cite{luca2}. Supersymmetric D3-branes should have vanishing D-term, which puts $e^{i \vartheta}=1$, indeed selecting
the warped Calabi-Yau vacua. These admit supersymmetric generalized four-cycles $(\Sigma,\calf)$, where $\Sigma$ must be a complex divisor inside $M$ and $\calf$ is restricted to be $(1,1)$ and primitive \cite{marchesano,luca1}. On these four-cycles instantons may wrap and generate non-perturbative corrections.

In this case the non-perturbatively corrected equation (\ref{gcscorr2}) implies that
\bea\label{npcorr}
\d\Z_{(1)}=-\frac{2i}{n} \calw_{\text{np}}\,\delta^{(2)}(\Sigma)\ .
\eea
Thus we see that, even if $\d\Z_{(1)}=0$ away from the locus of the instantonic cycle $\Sigma$, we cannot have in general $\Z_{(1)}\equiv 0$. This means that the ordinary complex structure is deformed into an honest generalized complex structure of type 1 implying in turn that a probe D3-brane feels a non-perturbatively generated superpotential \cite{luca2,paulluca}. This conclusion has already been reached from different points of view in \cite{ganor,haack,klebanov} and here we re-discover it from an alternative ten-dimensional geometrical interpretation.

To compare explicitly with the results of \cite{ganor,haack,klebanov} let us consider  $\Z_{(1)}$ as a small deformation of $\hat\Omega$, resulting into a small   deformation of the associated complex structure into a truly generalized complex one. If  $\Z_{(1)}$ is small, then to first approximation it must be a holomorphic $(1,0)$-form with respect  to the original background complex structure \cite{gualtieri}. Thus locally we can write $\Z_{(1)}=-(1/\pi)\partial\calw_{\text{D3}}$, where we are implicitly differentiating with respect to the position of the probe D3-brane. If we now consider not probe D3-branes, but D3-branes that are already present in our background, they should nevertheless feel the same non-trivial superpotential $\calw_{\text{D3}}$. On the other hand $\calw_{\text{np}}$ itself contains all the dependence on the space-filling D3-brane moduli\footnote{This dependence arises through the back-reaction of the D3-branes on the background fields which enter the non-perturbative correction \cite{ganor,klebanov}.}  \cite{ganor,klebanov} and we are thus led to identify $\calw_{\text{D3}}$ with $\calw_{\text{np}}$. More explicitly, putting $\calw_{\text{D3}}=\calw_{\text{np}}$ in eq.~(\ref{npcorr}) leads to
\bea\label{intequation}
\bar\partial(\partial \log \calw_{\text{np}})=\frac{2\pi i}{n}\,\delta^{(2)}(\Sigma)\ .
\eea
Since the dependence of $\calw_{\text{np}}$ on the D3-brane moduli is completely contained in $\cala$, (\ref{intequation}) becomes an equation for this prefactor, which can be readily integrated into
\bea\label{d3sup}
\cala=f^{1/n}\,\tilde\cala\ ,
\eea
where $\tilde\cala$ does not dependent on the D3-brane moduli and $f$ is the holomorphic section of the line bundle associated to the divisor $\Sigma$, which vanishes at the location of $\Sigma$ itself.\footnote{The vanishing of the superpotential at the location of the divisor comes from the open strings stretching from the D7/E3-brane and the space-filling D3-branes \cite{ganor,klebanov}.}

The resulting non-perturbative superpotential has thus the form
\bea\label{npd3}
\calw_{\text{np}}=f^{1/n}\tilde\calw_{\text{np}}\ ,
\eea
where $\tilde\calw_{\text{np}}$ does not depend on the space-filling D3-moduli, while this dependence is completely contained in $f$.
Note that by plugging (\ref{npd3}) back into $\Z_{(1)}=-(1/\pi)\d\calw_{\text{np}}$ we get, at the location of the D3-branes,
\bea
\Z_{(1)}=-\frac{1}{n\pi}f^{(1-n)/n}\d f\,\tilde\calw_{\text{np}}\ .
\eea
At weak coupling $g_s=\langle e^{\Phi}\rangle\ll 1$ so that $\tilde\calw_{\text{np}}$ is exponentially suppressed.  Then, if the non-perturbative correction is generated by instantonic D3-branes (i.e.~$n=1$), the small $\Z_{(1)}$ approximation is valid for any position of the D3-branes. On the other hand, when the non-perturbative effect  is induced by `confining'  D7-branes, the small $\Z_{(1)}$ approximation breaks down close to the D7-branes where $f=0$. Moreover the dilaton diverges so that $\tilde\calw_{\text{np}}$ is not guaranteed to be small either. This suggests that close to the  D7-branes the underlying Calabi-Yau complex structure seems not reliable anymore, so that we have to replaced it by a full generalized complex structure. Of course in this case classical IIB theory seems not reliable as well and it would be interesting to see if a similar effect can be recovered in an F-theory approach to the problem.

The final formula (\ref{d3sup}) is in perfect agreement with the results \cite{ganor,haack,klebanov} but gives a new ten-dimensional insight into  them. The key point is the deformation of the classical background complex structure into a non-perturbatively generated truly generalized complex structure, which is exactly what gives a non-trivial superpotential to the D3-branes in a geometric way along the lines of \cite{luca2,paulluca}.

\section{Conclusions}

In this paper we have shown that the ten-dimensional supersymmetry conditions for type II fully warped flux compactifications to AdS$_4$ or flat $\mathbb{R}^{1,3}$ space-time can be recovered from a purely four-dimensional supergravity description, albeit implicitly with an infinite number of fields. In particular, we have used a four-dimensional formulation that is invariant under local complexified Weyl transformations. This allows for a direct uplifting of the results to ten dimensions. By gauge-fixing the Weyl invariance one can obtain a standard Einstein-frame supergravity description \cite{supconf}, making contact with previous proposals in the literature to describe the four-dimensional theory corresponding to warped Calabi-Yau compactifications \cite{giddings1,giddings2}, and offering a natural interpretation and extension. Furthermore our results are consistent with the results for general $SU(3)$ or $SU(3)\times SU(3)$-structure backgrounds obtained in \cite{granaN2,grimm,granaN2bis}, in the
constant warp factor approximation, by dimensional reduction.

Our approach puts forward that a non-trivial warp factor, essential in many physically interesting scenarios, affects considerably the structure of the four-dimensional theory. For example, it explicitly breaks a possible underlying $N=2$ description to $N=1$ by entering the K\"ahler potential (\ref{classk}) in such a way as to couple to each other the bosonic fields that would be obtained from the vector- and hypermultiplets of the corresponding underlying $N=2$ theory in the constant warp factor approximation \cite{granaN2,granaN2bis}.  We do not present here a direct derivation of the full $N=1$ supergravity describing these warped compactifications, although it is important to investigate this more extensively. This should shed light on the question whether a low-energy truncated theory in which one consistently integrates out the higher Kaluza-Klein modes, leaving only massless or light modes, is still possible, as seems to be the case for unwarped compactifications assuming a large-volume limit. It should also allow to get a better handle on the inclusion of open string degrees of freedom corresponding to space-filling D-branes.  We leave these interesting problems for future research.

Finally, we have used our formulation to investigate, starting from the four-dimensional point of view, the effects on the internal geometry of possible non-perturbative corrections to the superpotential. Our arguments, even if relying on some unavoidable simplifying assumptions, suggest that corrections generated by D-brane instantons  or gauge theory strong coupling effects deform the generalized structure of the internal manifold. In particular, an ordinary complex structure can be seen as a special `unstable' point in the space of generalized complex structures on a given manifold \cite{gualtieri}. Thus, even if classically a supersymmetric vacuum is (possibly almost) complex in the ordinary sense, after perturbative and non-perturbative corrections are included, this property is expected to be lost. We have shown that these effects can be addressed somewhat more quantitatively in the case of IIB warped Calabi-Yau compactifications, which are equipped with the complex structure of the underlying Calabi-Yau manifold and can be affected by non-perturbative corrections arising from Euclidean D3-branes or stacks of confining D7-branes. These non-perturbative effects were crucial in \cite{kklt} to argue, from purely four-dimensional effective theory, for the existence of AdS vacua for these flux Calabi-Yau compactifications (in fact neglecting the non-trivial warp factor). We discuss how these AdS vacua can be understood from a ten-dimensional point of view by smearing the instantonic four-cycle generating the non-perturbative effect. Furthermore, we also considered more closely the effects of non-smeared instantons to the underlying ordinary complex structure. Neglecting additional terms related to the no-more vanishing cosmological constant, the net effect of an instanton is to produce a localized obstruction term for the integrability of the underlying complex structure. This in turn implies that it gets deformed to a truly generalized complex one, also producing a superpotential for mobile space-filling D3-branes \cite{luca2} in a geometric way. The resulting superpotential is in perfect agreement with the superpotentials computed in completely different ways in \cite{haack,klebanov}. In the case of confining D7-branes, our approximation seems to break down close to the internal divisor wrapped by the D7-branes and it would be very interesting to re-examine this effect in an F-theory approach. In any case, our results imply that generalized (complex) geometry should play an important role even for the mostly-used IIB warped Calabi-Yau or F-theory compactifications once possible perturbative or non-perturbative effects are taken into account.

\vspace{0.9cm}

\section*{Acknowledgements}
We are especially grateful to Toine Van Proeyen for illuminating explanations of superconformal supergravity and Thomas Grimm for collaboration
in the initial stage of the project. We also thank A.~Celi, G.~Dall'Agata, F.~Denef, M.~Haack, F.~Marchesano, J.~Rosseel, M.~Trigiante and M.~Zagermann for useful discussions.
L.~M.~acknowledges the Galileo Galilei Institute for Theoretical Physics for hospitality and the INFN for partial support.
Furthermore L.~M.\ is supported in part by the Federal Office for Scientific, Technical and Cultural Affairs through the ``Interuniversity Attraction Poles Programme -- Belgian Science Policy" P5/27 (2006) and P6/11-P (2007), and by the European Community's Human Potential Programme under contract MRTN-CT-2004-005104 `Constituents, fundamental forces and symmetries of the universe'. The work of P.~K.\ is supported by the
German Research Foundation (DFG) within the Emmy-Noether-Program (Grant number ZA 279/1-2).

\vspace{0.9cm}


\begin{appendix}

\section{Generalized geometry and Hodge decomposition}
\label{hodge}

From the supersymmetry of the background supergravity configuration follows that on the internal manifold $M$ there are two globally
defined polyforms $\Psi_1,\Psi_2 \in \Gamma(\Lambda^\bullet T^\star_M)$ that satisfy eqs.~(\ref{susyconds}) and \eqref{intconds}.
These polyforms can be regarded as global sections of the spinor bundle $\cals_{\pm}$ (of positive/negative chirality) associated to the generalized tangent bundle $T_M \oplus T^\star_M$. Indeed, a generalized vector $\mathbb{X}=(X,a) \in T_M \oplus T_M^\star$ acts on a polyform $\omega \in \Lambda^\bullet T^\star_M$ as
\eq{
\label{spinoraction}
\mathbb{X} \cdot \omega = \iota_X \omega + a \wedge \omega \, .
}
Because this action satisfies $\left(\mathbb{X}_1 \cdot \mathbb{X}_2 + \mathbb{X}_2 \cdot \mathbb{X}_1\right) \cdot \omega = 2 \, \cali(\mathbb{X},\mathbb{Y}) \omega $,
with the natural $(6,6)$-signature metric defined as
\eq{
\cali(\mathbb{X}_1,\mathbb{X}_2) = \frac{1}{2} \left( a_2(X_1) + a_1(X_2)\right) \, ,
}
it makes $T_M \oplus T_M^\star$ into a Clifford algebra and defines an isomorphism between the polyforms and the spinors of $T_M \oplus T_M^\star$.
The spaces of sections $\Gamma(\cals_+)$ and $\Gamma(\cals_-)$ correspond to the spaces of even and odd polyforms respectively. The Mukai pairing of two polyforms $\omega_1,\omega_2$ is defined as
\eq{
\label{mukai}
\langle \omega_1, \omega_2 \rangle = \omega_1 \wedge \alpha(\omega_2)|_{\text{top}} \, ,
}
where we select the top-form and $\alpha$ is the operator that reverses all the indices of a polyform. Under the equivalence with $T_M \oplus T_M^\star$-spinors
the Mukai pairing is identified with the spinor norm, from which we see that to properly define the isomorphism one needs to choose a volume form.

In the presence of an $H$-field it is natural to use -- as in for example eqs.~(\ref{susyconds}) and (\ref{intconds}) -- the $H$-twisted differential $\d_H=\d +H\wedge$ on globally defined polyforms. As an alternative, for any globally defined polyform $\omega$ one can consider the associated twisted polyform
\bea
\omega^\prime=e^B\wedge \omega\ ,
\eea
where $B$ is the NSNS 2-form potential, $\d B=H$, making the twisted polyform
in general only locally defined. The twisted polyforms can be seen as a global section of the spinor bundles $\cals'_{\pm}$ (of positive/negative chirality) associated to the extension bundle \cite{hitchin2}
\bea\label{ext}
0\rightarrow T_M^\star\rightarrow E\rightarrow T_M\rightarrow 0\ .
\eea
On the twisted polyforms the ordinary differential is the natural one, since $\d\omega^\prime=e^B\wedge\d_H\omega$. Twisted quantities are perhaps more natural to discuss the independent degrees of freedom since they also contain the $B$-field as we will see below. In the untwisted picture, the information on these degrees of freedom is contained in $H$. In fact, the
$B$-twist we have performed is a gauge transformation of the formalism that brings us from one extreme picture, where $B=0$ in the pure spinors to the other extreme picture where $H=0$. All expressions stay the same. The Mukai pairing, for instance, is invariant under the twist, i.e.\ $\langle\omega_1,\omega_2\rangle=\langle\omega^\prime_1,\omega^\prime_2\rangle$. If at all possible, we will use the untwisted globally defined polyforms.

Now, we define the null spaces $L_{1},L_{2} \subset T_M \oplus T^\star_M$ of $\Psi_1$ and $\Psi_2$ respectively,
i.e.\ $\mathbb{X} \in \Gamma(L_{1})$ if and only if $\mathbb{X} \cdot \Psi_1=0$ and analogously for $L_2$, and their
complex conjugates $\overline{L_{1}},\overline{L_{2}}$. $\Psi_1$ and $\Psi_2$ are pure, which means that $L_1$ and $L_2$
have the maximal dimension, in this case six. Moreover, since $\Psi_1$ and $\Psi_2$ are also compatible, \eqref{psconst2}, one can define the
three-dimensional intersections
\eq{
\begin{split}
L_1^+ & = L_1 \cap L_2 \, , \qquad L_1^- = L_1 \cap \overline{L_2} \, , \\
\end{split}
}
and complex conjugates. From these we can build the bundles $C_+,C_- \subset T_M \oplus T^\star_M$
defined as
\eq{
C_+ = L_1^+ \cup \overline{L_1^+} \, , \qquad
C_- = L_1^- \cup \overline{L_1^-} \, .
}
In general the elements of $C_+$ and $C_-$ have the form
\eq{
\label{genmetric}
\mathbb{X}_+ = (X,(\hat{h}+\hat{B})X) \in C_+ \, , \qquad \mathbb{X}_- = (X,(-\hat{h}+\hat{B})X) \in C_- \, ,
}
with $X \in T_M$. In the untwisted picture $\hat{h}=h$ is the internal metric and $\hat{B}=0$,
while repeating the above analysis for the twisted picture one sees that both $h=\hat{h}$ and $B=\hat{B}$ are contained in $\Psi'_1$
and $\Psi'_2$. Viewing the polyforms as ordinary spinor bilinears -- and thus in the untwisted picture -- the vectors of $C_+$ act as $SO(6)$ gamma-matrices on the left, while those of $C_-$ act on the right.

We can also define the generalized (almost) complex structures $\calj_1$ and $\calj_2$ associated to $\Psi_1$ respectively $\Psi_2$, which have as $+i$-eigenspaces
the bundles $L_1$ and $L_2$ respectively.

From the compatibility \eqref{psconst2} follows furthermore that we can construct the following generalized Hodge decomposition of differential forms, see \cite{gualtieri,gualtieri2}:
\begin{equation}\label{hodgedec}
\Lambda^\bullet \, T^\star_M \otimes \mathbb{C} = \bigoplus_{p,q} U_{p,q} \, ,
\end{equation}
where $U_{p,q}$ is the intersection of the $ip$-eigenspace of ${\cal J}_1$ and the $iq$-eigenspace of ${\cal J}_2$.  Complex conjugation sends thus $\overline{U_{p,q}} =U_{-p,-q}$. $\Psi_1$ and $\Psi_2$ are ``highest-state'' representations i.e.\ $\Psi_1\in\Gamma(U_{3,0})$ and $\Psi_2\in \Gamma(U_{0,3})$. Moreover, we
have the following behaviour for the action of elements of $L^\pm_1$ on $\omega_{p,q} \in U_{p,q}$
\eq{
\mathbb{X} \cdot \omega_{p,q} \in U_{p-1,q-1} \quad \text{for} \quad \mathbb{X} \in L^+_1 \, , \qquad
\mathbb{Y} \cdot \omega_{p,q} \in U_{p-1,q+1} \quad \text{for} \quad \mathbb{Y} \in L^-_1 \, ,
}
and complex conjugate relations. This includes the possibility that $\mathbb{X} \cdot \omega_{p,q} =0$ and the same for $\mathbb{Y}$.
It follows that one can form the elements of $U_{p,q}$ by acting with an antisymmetric product of $(3-(p+q))/2$ elements of $L_1^+$ and
$(3-(p-q))/2$ elements of $L_1^-$ on $\Psi_1$ or alternatively with an antisymmetric product of $(3-(p+q))/2$ elements of $L_1^+$
and $(3-(q-p))/2$ elements of $\overline{L_1^-}$ on $\Psi_2$.

Furthermore, the generalized Hodge decomposition is compatible with the Mukai pairing and the Hodge duality.
Indeed, if $\omega_{p,q}\in U_{p,q}$, then $\langle \omega_{p,q},\chi\rangle=0$ if $\chi$ is not in $U_{-p,-q}$, while Hodge duality
satisfies the following property
\begin{equation}
\label{hodgedual}
\omega_{p,q} = i(-1)^\frac{p+q-1}{2}\, \star_6 \!\left(\alpha(\omega_{p,q})\right) = \mp i (-1)^\frac{p+q-1}{2} \alpha(\star_6 \omega_{p,q}) \, ,
\end{equation}
for any $\omega_{p,q} \in U_{p,q}$ and $\mp$ corresponds to $\omega_{p,q}$ even/odd respectively. We will also sometimes use the partial decompositions with respect to the eigenvalues of only one of the generalized complex structures
\eq{
U_p = \bigoplus_{q} U_{p,q} \, , \qquad V_q = \bigoplus_{p} U_{p,q} \, .
}

For compactifications to flat Minkowski space, for which $W_0=0$, it follows from \eqref{susycond2} and \eqref{intcond2} that $\calj_2$ is integrable.
This implies that the exterior derivative $\d_H$ can only change the $q$-value by $\pm 1$, an in particular one can split $\d_H=\partial_H+\bar\partial_H$, where $\partial_H:\Gamma(V_{q})\rightarrow \Gamma(V_{q+1})$ and $\bar\partial_H:\Gamma(V_{q})\rightarrow \Gamma(V_{q-1})$. On the other hand,
for $\calj_1$, or also for $\calj_2$ if $W_0 \neq 0$, we do not have integrability, so that we cannot split $\d_H$ analogously. However, still the action of $\d_H$ on the $p$ or $q$-index in the decomposition (\ref{hodgedec}) is not completely unconstrained in terms of the Hodge decomposition. Indeed, from the identity
\bea
[\call^H_\mathbb{X},\mathbb{Y}\cdot]=[\mathbb{X},\mathbb{Y}]^C_H\cdot\ ,
\eea
where $[\cdot,\cdot]_H^C$ indicates the twisted Courant bracket and
\bea
\call^H_\mathbb{X}\equiv \d_H \mathbb{X}\cdot+\mathbb{X}\cdot \d_H\ ,
\eea
one can see, using an argument by induction, that
\bea\label{nonintdec}
\d_H: \Gamma(U_{p})\rightarrow \Gamma(U_{p-3})\oplus \Gamma(U_{p-1})\oplus\Gamma(U_{p+1})\oplus\Gamma(U_{p+3})\ ,
\eea
and analogously for the $V_q$-decomposition.

\section{Deformations of $\T$ and $\Z$}
\label{infdef}

In this section we want to describe in some detail the holomorphic structure of the field space defined by $\Z^\prime$ and $\T^\prime$. In particular we want to identify a set of independent holomorphic infinitesimal deformations of $\Z^\prime$ and $\T^\prime$. As in the rest of the paper we will go to the untwisted picture.
This is possible since even if the untwisted $\T$ and $\Z$ do not contain information on $B$, we can still consider all deformations as fluctuations of $\T$ and $\Z$, including deformations of the $B$-field. To stay in the untwisted picture one would in a second step absorb the latter deformations in $H$ by making a compensating small twist over $-\delta B$. The discussion would essentially be identically in the twisted picture anyway, simply adding primes everywhere and keeping in the back of our minds that the Hodge decomposition is then
with respect the twisted pure spinors.

Let us start by considering the pure spinor $\Z$ and the stable spinor $\Re \T=\Re \t$. Since $\Re \t$ determines $\Im \t$ as explained in \cite{hitchin}, together they
define the $SU(3)\times SU(3)$-structure (which implies, once twisted, the metric and the $B$-field), the dilaton and the warp factor. The $SU(3)\times SU(3)$-structure constraint can be written as
\bea\label{appconst}
\langle \Z,\mathbb{X}\cdot \Re \T\rangle=0 \, ,\qquad \forall \mathbb{X}\in T_M\oplus T^\star_M\ .
\eea
In terms of the Hodge decomposition of appendix \ref{hodge} the most general deformation of $\Z$ satisfies $\delta \Z \in \Gamma(V_3 \oplus V_1)$,
where the part in $\Gamma(V_1)$ describes the deformations of the associated generalized complex structure $\calj_2$ \cite{gualtieri} and the part
in $\Gamma(V_3)$ affects the warp factor.
It can be further split as follows
\bea
\delta\Z=\delta_{1}\Z +\delta_3\Z+\delta_4\Z \, ,
\eea
with
\bea\label{decz}
\delta_1\Z\in\Gamma(U_{0,3}\oplus U_{0,1}) \, , \qquad \delta_3\Z\in\Gamma(U_{-2,1})\quad\text{and}\quad \delta_4\Z\in\Gamma(U_{2,1})\ .
\eea
Similarly we have $\delta \t \in \Gamma(U_3 \oplus U_1)$, where the part in $\Gamma(U_1)$ describes the deformations
of the other generalized complex structure $\calj_1$ and this time the part in $\Gamma(U_3)$ affects both the dilaton and the warp
factor. It follows that the most general deformation of $\Re\T$ has the form
\bea
\delta\Re\T=\delta_2\Re\T +\delta_3\Re\T+\delta_4\Re\T\ ,
\eea
with
\bea\label{str1}
&\delta_2\Re\T\in\Gamma(U_{3,0}\oplus U_{1,0}\oplus U_{-1,0}\oplus U_{-3,0})\, ,\qquad \delta_3\Re\T\in\Gamma(U_{1,-2}\oplus U_{-1,2})&\quad\cr &\text{and}\quad \delta_4\Re\T\in\Gamma(U_{1,2}\oplus U_{-1,-2})\ .&
\eea
Of course, since $\Re\T$ is real we have
\bea\label{rc}
\delta_2\Re\T|_{U_{-3,0}}=\overline{\delta_2\Re\T|_{U_{3,0}}}\, ,\qquad \delta_2\Re\T|_{U_{-1,0}}=\overline{\delta_2\Re\T|_{U_{1,0}}}\, ,\quad \text{etc.}
\eea
Now, $\delta\T$ contains also the RR variations $\delta C$, which can be decomposed in the same way as we did for $\delta\Re\T$
\bea\label{str2}
&\delta_2 C\in\Gamma(U_{3,0}\oplus U_{1,0}\oplus U_{-1,0}\oplus U_{-3,0})\, , \qquad\delta_3 C\in\Gamma(U_{1,-2}\oplus U_{-1,2})&\quad\cr &\text{and}\quad \delta_4C\in\Gamma(U_{1,2}\oplus U_{-1,-2})\ ,&
\eea
with obvious reality conditions similar to (\ref{rc}).

Let us now impose the constraint (\ref{appconst}). It is easy to see that it is automatic for the deformations $\delta_1\Z$ and $\delta_2\Re\T$.
These deformations, to be precise the parts in $\Gamma(U_1)$ and $\Gamma(V_1)$, were already found in \cite{gualtieri} as the deformations of
the $SU(3)\times SU(3)$-structure that affect one generalized complex structure without touching the other. It also turns out they are exactly
the ones that deform the generalized metric $(h,B)$.
Furthermore, the RR deformations are unconstrained, so that we can write out a first set of independent holomorphic transformations
\bea
\delta_1\Z\in\Gamma(U_{0,3}\oplus U_{0,1})\quad\text{and}\quad \delta_2\T\in\Gamma(U_{3,0}\oplus U_{1,0}\oplus U_{-1,0}\oplus U_{-3,0})\ ,
\eea
where now the different components of $\delta_2\T$ are unrelated since $\T$ is complex.

On the other hand the constraint (\ref{appconst}) relates  the deformations $\delta_3$ in (\ref{decz}) and (\ref{str1}) and the
same for $\delta_4$. Hence the labelling. While for definiteness we focus on $\delta_3$, the discussion is completely analogous for $\delta_4$. It is easy to see that any deformation $\delta_3\Z$ must be accompanied by a corresponding deformation $\delta_3\Re\T$ and viceversa. In more detail, we can parameterize such a deformation by a complex $\varepsilon\in \Gamma(\Lambda^2 \overline{L^+_1})$, so that
\bea
\delta_3\Z=\varepsilon\cdot \Z\in\Gamma(U_{-2,1})\quad\text{and}\quad \delta_3\Re\T=\Re(\varepsilon\cdot t)\in\Gamma(U_{1,-2}\oplus U_{-1,2})\ .
\eea
Note that we can also use an analogous complex parameter $\varepsilon^\prime\in \Gamma(\Lambda^2 \overline{L^+_1})$ to define a general $\delta_3$ deformation of the RR fields
\bea
-\delta_3 C=\delta_3\Im\T=\Re(\varepsilon^\prime\cdot t)\in\Gamma(U_{1,-2}\oplus U_{-1,2})\ .
\eea
So we see that the $\delta_3$ deformations respecting the constraint (\ref{appconst}) are parameterized by the complex parameters $\varepsilon$ and $\varepsilon^\prime$, in terms of which $\delta_3\T$ has the form
\bea
\delta_3\T=(\varepsilon+i\varepsilon^\prime)\cdot t+ (\bar\varepsilon+i\bar\varepsilon^\prime)\cdot \bar t\ .
\eea
This means that $\varepsilon$ and $\varepsilon^\prime$ can in turn be expressed in terms of $\delta_3\Z$ and $\delta_3\T|_{U_{1,-2}}$, which
we can thus consider as independent holomorphic deformations. The same argument can be repeated for the $\delta_4$ deformations, which can be holomorphically parameterized by $\delta_4\Z$ and $\delta_4\T|_{U_{-1,-2}}$. We remark also that the deformations $\delta_3,\delta_4$ do not deform the spaces $C_+$ and $C_-$, and
hence do not affect the generalized metric $(h,B)$.

The final outcome of this discussion is that we can take as independent holomorphic deformations of $\Z$ and $\T$ the following
\bea
\label{ddecomp}
\delta\T\in\Gamma(V_{0}\oplus V_{-2})\quad \text{and} \quad \delta\Z\in\Gamma(V_{3}\oplus V_{1})\ .
\eea
Note that the holomorphic fluctuations (\ref{ddecomp}) can be identified by using the Hodge decomposition associated to $\calz$ alone, without using any other structure. The complex structure they define is compatible with the natural one for $\Z$ but not with the natural one for $\T$, since both $\delta\T|_{U_{\pm 1,2}}$ are anti-holomorphic functions of both $\delta\Z|_{U_{\pm 2,1}}$ and $\delta\T|_{U_{\pm 1,-2}}$.

\section{Weyl invariant $N=1$ supergravity}
\label{sugrapp}

We start from the superconformal supergravity discussed in \cite{supconf}. Let us indicate the chiral fields with $\phi^I$ and use conventions similar to \cite{supconf} for writing complex conjugated quantities, e.g.\ $\bar\phi_I\equiv (\phi^I)^*$, and derivatives e.g.\ $\caln_I=\partial\caln/\partial\phi^I$. Omitting scalar and vector kinetic terms, the bosonic Lagrangian has the form
\bea
(-\det g)^{-1/2}\call=\frac12 \, \caln \, R+\frac13 \, \calw^{*I}(\caln^{-1})_{I}{}^J\calw_J-\frac12 \, (\Re f)^{-1}(\cald,\cald)+\ldots
\eea
where $\caln$ is a real function of $\phi^I$ and $\bar\phi_I$, $\calw$ depends holomorphically on the $\phi^I$s and is related to the standard superpotential as in eq.~(\ref{einsup}), $\Re f$ is the metric of the vector multiplets -- which is the real part of a holomorphic function of the $\phi^I$s -- and finally $\cald$ represents the D-terms (or Killing potentials), which symplectically generate the gauged isometries (see eq.~(\ref{dterm})).

The different chiral fields transform with certain weights $(w,c)$ under the dilatation and chiral transformations\footnote{Note that our convention for the four-dimensional chiral operator is opposite to the one of \cite{supconf}, so that chiral here corresponds to anti-chiral there.}
\bea\label{cc}
\phi\rightarrow e^{w\sigma+ic\alpha}\phi\ ,
\eea
and $\caln$ and $\calw$ must transform homogeneously under dilatation and chiral transformations with weights $(2,0)$ and $(3,1)$ respectively. For example, in our setting $\Z$ has weights $(3,1)$ while $\T$ has weights $(0,0)$.

For simplicity, in this appendix we assume that the chiral fields $\phi^I$ have been redefined such that they all have weights $(1,1/3)$. Then, for our purposes, the relevant terms in the superconformal transformations of the gravitino $\psi_\mu$, the fermions of the chiral multiplets $\chi_I$ and the gauginos $\lambda$ are given by
\eq{\label{ss}
\begin{split}
\delta\psi_\mu & = \nabla_\mu\zeta_- -\gamma_\mu\xi_+ +\ldots\, , \\
\delta\chi_I & = \frac16\calw_{J}(\caln^{-1})^{J}{}_I\zeta_+ + \bar\phi_I\xi_+ +\ldots \, , \\
\delta\lambda & = -\frac{i}{2}(\Re f)^{-1}\cald\zeta_++\ldots\, ,
\end{split}
}
where $\zeta_+$ and $\xi_+$ are the generators of the standard ($Q$-)supersymmetry and the $S$-supersymmetry respectively.

To partially break the superconformal invariance while keeping the invariance under the bosonic gauge symmetry (\ref{cc}) it is sufficient to  eliminate the gauge field of the dilatations (which we have not mentioned at all and is called $b_\mu$ in \cite{supconf}), and gauge-fix the $S$-supersymmetry by eliminating one of the spinors of the chiral multiplets (the one associated to the compensator). This can be explicitly done by imposing the gauge-fixing \cite{kugo}
\bea\label{Sgauge}
\caln^I \, \chi_I=0\ .
\eea
This fixes $\xi_+$ uniquely as
\bea
\xi_+=-\frac{\calw}{2\caln}\zeta_+\ ,
\eea
so that after the gauge-fixing the standard supersymmetry transformations take the form
\eq{\label{finsusy}
\begin{split}
\delta\psi_\mu & = \nabla_\mu\zeta_- +\frac{\calw}{2\,\caln}\gamma_\mu\zeta_+ +\ldots \, , \\
\caln_{I}{}^J\delta\chi_J & = \frac16\Big(\calw_{I} - \frac{3\,\caln_I}{\caln}\,\calw\Big)\zeta_+ +\ldots \, , \\
\delta\lambda & = -\frac{i}{2}(\Re f)^{-1}\cald\zeta_++\ldots \, .
\end{split}
}


\section{Supersymmetric D-brane instantons in flux vacua}
\label{appinst}

In this appendix we will introduce supersymmetric D-brane instantons, or shortly E-branes, and derive their calibration form. E-branes live on an Euclidean background
which we obtain by analytical continuation from the Minkowskian one.

We take the ten-dimensional gamma matrices to split as follows into four- and six-dimensional gamma-matrices
\bea
\Gamma^\mu=\gamma^\mu\otimes \bbone\, , \qquad \Gamma^m=\gamma_{(4)}\otimes \hat\gamma^m\ ,
\eea
 where $\gamma_{(4)}=i\gamma^{\ul{0123}}$ is the four-dimensional chiral operator, while the six-dimensional internal one is given by $\hat\gamma_{(6)}=-i\hat\gamma^{\ul{123456}}$.
Under a Wick rotation $x^0\rightarrow -ix^0$ we have to correspondingly rotate $\Gamma^{\ul{0}}\rightarrow -i\Gamma^{\ul{0}}$. Thus the ten-dimensional gamma matrices are no longer real and we must relax the reality condition on $\epsilon_{1,2}$. This can be achieved by relaxing in turn the reality condition on $\zeta$ and considering $\zeta_+$ and $\zeta_-$ in \eqref{adskilling} as independent spinors which are chiral with respect to the (Euclidean) 4d chiral operator $\gamma_{(4)}=\gamma^{\ul{0123}}$. This procedure must be seen as an analytical continuation, where the total number of independent (holomorphic) components contained in $\zeta_+$ and $\zeta_-$ is still four, like in the Minkowskian case.

Let us now take an E-brane stretching in $p+1$ dimensions, which we will call an E$p$-brane -- following the naming conventions of D-branes as is customary -- that wraps a Euclidean generalized cycle $(\Sigma,\calf)$ inside $M$. The E$p$-brane bosonic action $S_{\text{E}}$ in the Wick-rotated vacuum  is given by
\bea\label{Eaction}
S_{\text{E}}=2 \pi \int_\Sigma \d^{p+1}\sigma\, e^{-\Phi}\sqrt{\det (g|_\Sigma+\calf)}-2 \pi i\int_\Sigma C|_\Sigma\wedge e^\calf\ ,
\eea
and enters the path-integral via $\exp(-S_{\text{E}})$. The fermionic completion of (\ref{Eaction}) can be described in the Green-Schwarz formalism by considering two ten-dimensional Weyl spinors $\theta_1$ and $\theta_2$ as world-volume dynamical fields. Furthermore the resulting action has a gauge $\kappa$-symmetry that around a bosonic configuration takes the form
\bea\label{kappa}
\delta_\kappa\theta_1=\kappa\, , \qquad \delta_\kappa\theta_2=\Gamma_{\text{E}}^{-1}\kappa\ ,
\eea
where $\kappa$ is a Weyl spinor of positive chirality and
\bea
\Gamma_{\text{E}}=\frac{i}{\sqrt{\det(g|_\Sigma+\calf)}}\sum_{2l+s=p+1}\frac{\epsilon^{\alpha_1\ldots\alpha_{2l}\beta_1\ldots\beta_s}}{l!s!2^l}\calf_{\alpha_1\alpha_2}\cdots\calf_{\alpha_{2l-1}\alpha_{2l}}\Gamma_{\beta_1\ldots\beta_s}\ .
\eea

From the $\kappa$-symmetry (\ref{kappa}),  an instantonic E$p$-brane preserves  a Killing spinor $(\epsilon_1,\epsilon_2)$  if and only if
\bea\label{susycond}
\epsilon_1=\Gamma_{\text{E}}\epsilon_2\ .
\eea
Since in the Euclidean frame, $\zeta_+$ and $\zeta_-$ are independent,
we have two kinds of background Killing spinors
\bea
\quad\text{chiral:}&&\quad \epsilon_1^{{L}}=\zeta_+\otimes\eta^{(1)}_+\, , \qquad \epsilon_2^{{L}}=\zeta_+\otimes\eta^{(2)}_\mp\ ,\cr
\quad\text{anti-chiral:}&&\quad \epsilon_1^{{R}}=\zeta_-\otimes\eta^{(1)}_-\, , \qquad \epsilon_2^{{R}}=\zeta_-\otimes\eta^{(2)}_\pm\ .
\eea
By imposing (\ref{susycond}) for $(\epsilon_1^{{L}},\epsilon_2^{{L}})$ one gets
\bea
\hat\gamma^\prime_{(p+1)}\eta^{(2)}_\mp=-i\eta^{(1)}_+\ ,
\eea
for IIA/IIB respectively, and
\bea
\hat\gamma^\prime_{(p+1)}=\frac{1}{\sqrt{\det(g|_\Sigma+\calf)}}\sum_{2l+s=p+1}\frac{\epsilon^{\alpha_1\ldots\alpha_{2l}\beta_1\ldots\beta_s}}{l!s!2^l}\calf_{\alpha_1\alpha_2}\cdots\calf_{\alpha_{2l-1}\alpha_{2l}}\hat\gamma_{\beta_1\ldots\beta_s}\ .
\eea
On the other hand, imposing  the supersymmetry preservation on  $(\epsilon_1^{{R}},\epsilon_2^{{R}})$ one gets
\bea
\hat\gamma^\prime_{(p+1)}\eta^{(2)}_\mp=i\eta^{(1)}_+\ ,
\eea
again for IIA/IIB respectively.

We thus see that an instantonic E-brane can preserve only half of the background supersymmetry.  We will refer to E-branes preserving the chiral Killing spinors  $(\epsilon_1^{{L}},\epsilon_2^{{L}})$  as {\em instantonic} E-branes, while to those preserving the anti-chiral Killing spinors  $(\epsilon_1^{{R}},\epsilon_2^{{R}})$  as {\em anti-instantonic} E-branes.
Comparing with the results of \cite{luca1}, we see that supersymmetric instantonic E-branes wrap exactly the same generalized calibrated cycles in the internal space as supersymmetric space-time filling branes, i.e.\ they satisfy the condition
\bea\label{calcond}
\sqrt{\det(g|_\Sigma+\calf)}=\left.\Re\Psi_1|_\Sigma\wedge e^\calf\right|_{\text{top}}\ ,
\eea
while anti-instantonic  E-branes have opposite orientation.

One can thus borrow several   results for space-time filling D-branes discussed in \cite{luca1,luca2,paulluca}. In particular, one can split the supersymmetry/calibration condition (\ref{calcond}) in a pair of conditions that, in the case of space-filling D-branes, correspond to the F-flatness and the D-flatness of the four-dimensional description \cite{luca1,luca2}. The F-flatness requires $(\Sigma,\calf)$ to be an (almost) generalized complex cycle. Using the dual generalized current, it can be rephrased as  $j_{(\Sigma,\calf)} \in \Gamma(V_0)$ \cite{paulluca}. For example,  IIB $SU(3)$-structure compactifications are complex and in this case the F-flatness requires $\Sigma$ to be  holomorphically embedded and $\calf^{2,0}=0$. The D-flatness condition requires,  in the notation of this paper, that $\langle\Im t,j_{(\Sigma,\calf)}\rangle=0$. For example, in the case of a four-cycle on a IIB warped Calabi-Yau compactification, it requires the primitivity of $\calf$ with respect to the underlying K\"ahler structure (see \cite{luca1,luca2,paulluca} for more details and examples).

We proceed with the instantons and find that the action \eqref{Eaction}
reduces to
\bea\label{instaction}
S_{\text{E}}(\T)=2 \pi \int_\Sigma ( e^{-\Phi} \Re \Psi_1 - i C)|_\Sigma \wedge e^\calf = 2 \pi \int_M \langle \T, j_{(\Sigma,\calf)} \rangle \, ,
\eea
pointing to the definition \eqref{hol2} of $\T$ as a natural holomorphic field.
The same identification can be obtained from space-filling D-branes. Indeed, using the results of \cite{luca1,luca2} it is easy to see that the four-dimensional bosonic action for the $U(1)$ gauge fields living on the dimensionally reduced  D-brane is given by
\bea
\label{gaugeF}
-\frac{1}{16\pi^2}\int_X\d^4x\sqrt{-\det g} \, \Re S_{\text{E}}(\T) \, {\rm F}_{\mu\nu} {\rm F}^{\mu\nu} -\frac{1}{8\pi^2}\int_X \Im S_{\text{E}}(\T) \, {\rm F}\wedge {\rm F} +\ldots
\eea
where ${\rm F}$ is the field-strength of the massless gauge field arising in the dimensional reduction (in our units $\calf_{\mu\nu} = \text{F}_{\mu\nu}/2\pi$). The ellipsis stands for higher Kaluza-Klein contributions that can be  simply obtained by considering a more general $\text{F}$, which depends also on the internal world-volume coordinates, and `moving' it inside the integrals in (\ref{instaction}).


As a final observation, let us recall that a general  infinitesimal deformation of a generalized cycle is described by a section $\mathbb{X}$ of the generalized normal bundle $\caln_{(\Sigma,\calf)}$ \cite{luca2}. From the analysis of \cite{paulluca} the variation of the action under such a deformation is given by
\eq{
\delta_{\mathbb{X}} S_{\text{E}} = 2 \pi \int_M \langle \d_H\T,\mathbb{X} \cdot j_{(\Sigma,\calf)} \rangle \, .
}
It follows immediately from \eqref{susy21} that the instanton action is invariant under variations $\mathbb{X} \in \Gamma(\caln^{(1,0)}_{(\Sigma,\calf)})$ with $\caln^{(1,0)}_{(\Sigma,\calf)}=\caln_{(\Sigma,\calf)} \cap L_2$. In \cite{paulluca} the sections of $\Gamma(\caln^{(0,1)}_{(\Sigma,\calf)})=\caln_{(\Sigma,\calf)} \cap \bar L_2$ were used to parameterize the holomorphic deformations of generalized cycles and thus we see how the instanton action (\ref{instaction}) depends holomorphically not only on the closed string degrees of freedom but also on the open string ones. Similarly, the anti-instanton action is invariant under the variations $\mathbb{X} \in \caln^{0,1}_{(\Sigma,\calf)}$ and thus it is anti-holomorphic in both the closed and open string degrees of freedom.

\section{Orientifolds in generalized flux compactifications}
\label{orientifolds}

A proper flux compactification usually needs the inclusion of orientifolds to satisfy tadpole conditions and circumvent no-go theorems \cite{nogo}. Orientifolds also break  supersymmetry explicitly (off-shell), reducing in our case a possible underlying $N=2$ supergravity description to $N=1$. Orientifolds in $SU(3)\times SU(3)$ compactifications were studied before in e.g.~\cite{grimm,scan,dimi}.

An orientifold action $\calo$ is a composition of a reflection on the world-sheet
(denoted by $\Omega_p$) exchanging the left-movers with the right-movers,
and a target-space involution $\sigma$ ($\sigma^2=1$ on bosonic fields) acting
on the internal manifold. A factor $(-1)^{F_L}$, where $F_L$ is the
fermion number of the left-movers, is sometimes needed to ensure $\calo^2=\bbone$
on all states including spinors. Whether it appears or not depends on the
number of $+1$-eigenvalues of $\sigma$, which also determines the dimensionality of
the orientifold plane. This is the fixed point set of the involution which,
in our case, fills the four-dimensional space-time.

For the dilaton $\Phi$, metric $h$ and NSNS three-form $H$ to be invariant under the total orientifold projection $\calo$, they have to transform under the involution as
\begin{equation}\label{nsproj}
\sigma^* \Phi = \Phi \, , \qquad \sigma^* h = h \, , \qquad \sigma^*H = -H \, ,
\end{equation}
while for the RR potentials we need
\eq{
\label{sigmaRR}
\sigma^* C = \alpha(C) \quad (O3/O4/O7/O8) \, , \qquad \sigma^* C  = \alpha(C) \quad (O5/O6/O9) \, ,
}
For $N=1$ supersymmetric orientifolds we need to have moreover for the pure spinors
\begin{subequations}
\label{sigmapurespinors}
\begin{align}
& & \sigma^* \Psi_1 & = \alpha(\bar{\Psi}_1) \quad (O3/O4/O7/O8) \, , & \sigma^* \Psi_1 & = - \alpha(\bar{\Psi}_1) \quad (O5/O6/O9) \, ,  \\
& & \sigma^* \Psi_2 & = \alpha(\Psi_2) \quad (O3/O6/O7) \, , & \sigma^* \Psi_2 & = - \alpha(\Psi_2) \quad (O4/O5/O8/O9) \, .
\end{align}
\end{subequations}
It follows that for the holomorphic variables
\begin{subequations}\label{holproj}
\label{sigmaholo}
\begin{align}
& & \sigma^* \T & = \alpha(\T) \quad (O3/O4/O7/O8) \, , & \sigma^* \T & = - \alpha(\T) \quad (O5/O6/O9) \, ,  \\
& & \sigma^* \Z & = \alpha(\Z) \quad (O3/O6/O7) \, , & \sigma^* \Z & = - \alpha(\Z) \quad (O4/O5/O8/O9) \, .
\end{align}
\end{subequations}

The same conditions (\ref{holproj}) are equally valid for the twisted variables $\T^\prime$ and $\Z^\prime$ since from (\ref{nsproj}) we have $\sigma^*B = -B$.
One can easily check that the various Mukai pairings, appearing in the integrand of the superpotential \eqref{classup} and the K\"ahler potential \eqref{classk},
transform under the orientifold involution appropriately as the volume form, which means they change sign in type IIA and stay invariant in type IIB. Furthermore,
we will always implicitly assume that we integrate over the covering space of the orientifold action and then divide by an appropriate factor to avoid overcounting.
\end{appendix}


\end{document}